# Emissions Assessment of Low Earth Orbit (LEO) Broadband Megaconstellations; Starlink, OneWeb and Kuiper.

OGUTU B. OSORO[1,2], EDWARD J OUGHTON[2], ANDREW WILSON[3] and AKHIL RAO[4]

[1] Department of Geography & GIS, University of Cincinnati, Cincinnati, OH 45221, USA.

[2] Department of Geography & Geoinformation Science Department, George Mason University, Fairfax, VA 22030, USA

[3] Department of Construction & Built Environment, School of Science & Engineering, Glasgow Caledonian University, Glasgow, G4 0BA, UK.

[4] Department of Economics, Middlebury College, Middlebury, VT 05753, USA.

## Abstract

The growth of Low Earth Orbit (LEO) broadband satellite megaconstellations is rapidly increasing the number of rocket launches. While improving broadband Internet helps achieve the Sustainable Development Goals (SDGs), there are also significant environmental emissions produced from burning rocket fuels. We present sustainability analytics for *phase 1* of the three main LEO constellations including Amazon Kuiper (3,236 satellites), Eutelsat Group's OneWeb (648 satellites), and SpaceX Starlink (4,425 satellites). We find that LEO megaconstellations provide substantially improved broadband speeds for rural and remote communities but are roughly 6-8 times more emissions intensive (250 kg $CO_2$eq/subscriber/year) than comparative terrestrial 4G mobile broadband. Policy makers must carefully consider the trade-off between improving broadband Internet to further the SDGs while mitigating the growing space sector environmental footprint, particularly regarding *phase 2* plans to launch an order-of-magnitude more satellites.

## 1. Introduction

Low Earth Orbit (LEO) broadband megaconstellations involve plans to launch tens of thousands of new satellites into space to provide global broadband Internet coverage. Indeed, LEO constellations are a key reason why the quantity of rocket launches to space has rapidly increased, from fewer than 250 launches annually in the 1970s (below 2,900 per decade), to now exceed 1,300 launches annually (~16,000 in a single decade, more than a 400% increase)[1]. This growing number has resulted in a range of emerging questions around the negative environmental externalities of these satellite megaconstellations (Witze, 2023) and the environmental sustainability aspects of this approach(Lawrence et al., 2022; Miraux, 2022a), given the increasing commercialization of space activities, from tourism to earth observation. For instance, past studies have demonstrated that anthropogenic Greenhouse Gases (GHGs) such as oxides of nitrogen and carbon are responsible for cooling of the upper atmosphere (Keating et al., 2000). At the same time, these GHGs cause the warming of the lower atmosphere thus creating a vertical temperature gradient that might disrupt established weather patterns (Santer et al., 2023). Additional studies have further established that the deposition of GHGs in the upper atmosphere from rocket launches reduces the atmospheric density, thus decreasing the efficiency of debris removal (Lewis et al., 2011).

Indeed, the shift towards satellite megaconstellations raises new environmental sustainability questions around rocket launch emissions (Dallas et al., 2020), with evidence-based studies indicating projected trends are likely to produce adverse impacts, attracting the attention of regulatory authorities (Miraux, 2022b). It is therefore imperative that governments carefully balance the growth of the space sector, and the associated benefits in progressing the Sustainable Development Goals (SDGs), against environmental sustainability issues (Wilson et al., 2022). This provides strong research motivation for assessing the sustainability implications of launching the large numbers of planned LEO satellites in key megaconstellations. Particularly as the space sector is growing at such a rapid rate as to concern those outside the space community. Evidence is urgently needed to understand the negative environmental impacts of rocket launch emissions, to direct mitigation strategies, as evaluated here.

Importantly, "what goes up, must come down". Hence, there are also emerging concerns around re-entry emissions. Currently, regulators have strict requirements for LEO satellites to be decommissioned by either re-entry or moving assets to higher orbits. The re-entry of satellites results in emissions that could also have a substantial climate and ozone impact (Ryan et al., 2022). Recent estimates suggest satellite and rocket re-entry is responsible for 10% of aerosols in the stratosphere (Murphy et al., 2023), potentially depositing more alumina on the upper atmosphere than meteoroids (Boley and Byers, 2021). At this stage, the science of re-entry impacts is still emerging and is an active area of future research.

In this paper we develop an integrated model capable of assessing the environmental impacts associated with rocket launches for specific phase 1 LEO constellations, with concurrent metrics on the provided capacity, financial costs and Social Cost of Carbon (SCC). We treat phase 1 of each LEO constellation as the filing information submitted to the US Federal Communications Commission (FCC), such as for Amazon's Kuiper (3,236 satellites), Eutelsat Group's OneWeb (648 satellites) and SpaceX's Starlink (4,425 satellites)(Federal Communications Commission, 2020a, 2018a, 2018b). Additionally, a representative

[1] Our World in Data, 2023. Number of payloads and rocket bodies in space, by orbit. URL https://ourworldindata.org/grapher/space-objects-by-orbit (accessed 5.31.23).

Geostationary Earth Orbit (GEO) constellation is appraised for comparison. Currently, within the GEO satellite industry there are a number of major operators such as Intelsat (52 satellites)[2], Eutelsat (35 satellites)[3], Inmarsat (14 satellites)[4], Arabsat (8 satellites)[5], ViaSat (4 satellites)[6] and Avanti (4 satellites)[7]. Here we utilize a representative GEO operator treated as having 19 satellites, representing the mean quantity across these major operators.

Policy makers must consider a key trade-off regarding the SDGs. On the one hand, the delivery of broadband Internet to unconnected communities is recognized to progress the SDGs. While on the other, the results of this paper demonstrate that the rapid growth in the satellite sector is a pressing issue with substantial environmental sustainability implications. The method and evidence presented (i) informs future space sustainability metrics such as the Space Sustainability Rating (SSR) system, and (ii) supports strategic future choices in rocket design and fuel options.

## 2. Literature Review

In very remote areas, where terrestrial broadband infrastructure is not economically viable due to low population density and/or low adoption, LEO satellites provide high-capacity, low-latency broadband Internet to hard-to-reach communities (Ahmmed et al., 2022; del Portillo et al., 2021). Importantly, there are key differences when compared to traditional GEO constellations. For example, LEO satellites are designed to be smaller in size, with shorter lifespans (e.g., 5 years), and are therefore less costly to produce (Osoro and Oughton, 2021). However, the more proximate orbit to Earth means many more satellites are required to achieve global coverage (Maral et al., 2020).

That has prompted satellite operators to file for a very large number of satellites in their proposed constellation designs, with the expectation that each constellation will need to be continuously replenished as older satellites end their operational life. For example, in 2022 there were ~6,300 satellites in operation, whereas the total number of proposed satellites over the next decade in new constellations will be as high as ~320,000 (Federal Communications Commission, 2018b). Fig. 1 shows the technical details of three operational and planned LEO constellations assessed in this paper.

[2] Intelsat Ltd, 2024. Satellite Flight Operations and Engineering. URL https://www.intelsat.com/space/products/satellite-operations/ (accessed 1.15.24).

[3] Eutelsat, 2024. Our Company. URL https://www.eutelsat.com (accessed 2.19.24).

[4] Inmarsat Corp, 2024. Global coverage. Infinite connections. Website. URL https://www.inmarsat.com/en/about/technology/satellites.html (accessed 2.19.24).

[5] Arabsat, 2024. The Arabsat Fleet. URL https://www.arabsat.com/the-fleet/ (accessed 2.19.24).

[6] Viasat, 2023. An entire world of customers relies on Viasat satellites. URL https://news.viasat.com/blog/corporate/an-entire-world-of-customers-relies-on-viasat-satellites (accessed 2.19.24).

[7] Eutelsat, 2024. Our Company. URL https://www.eutelsat.com (accessed 2.19.24).

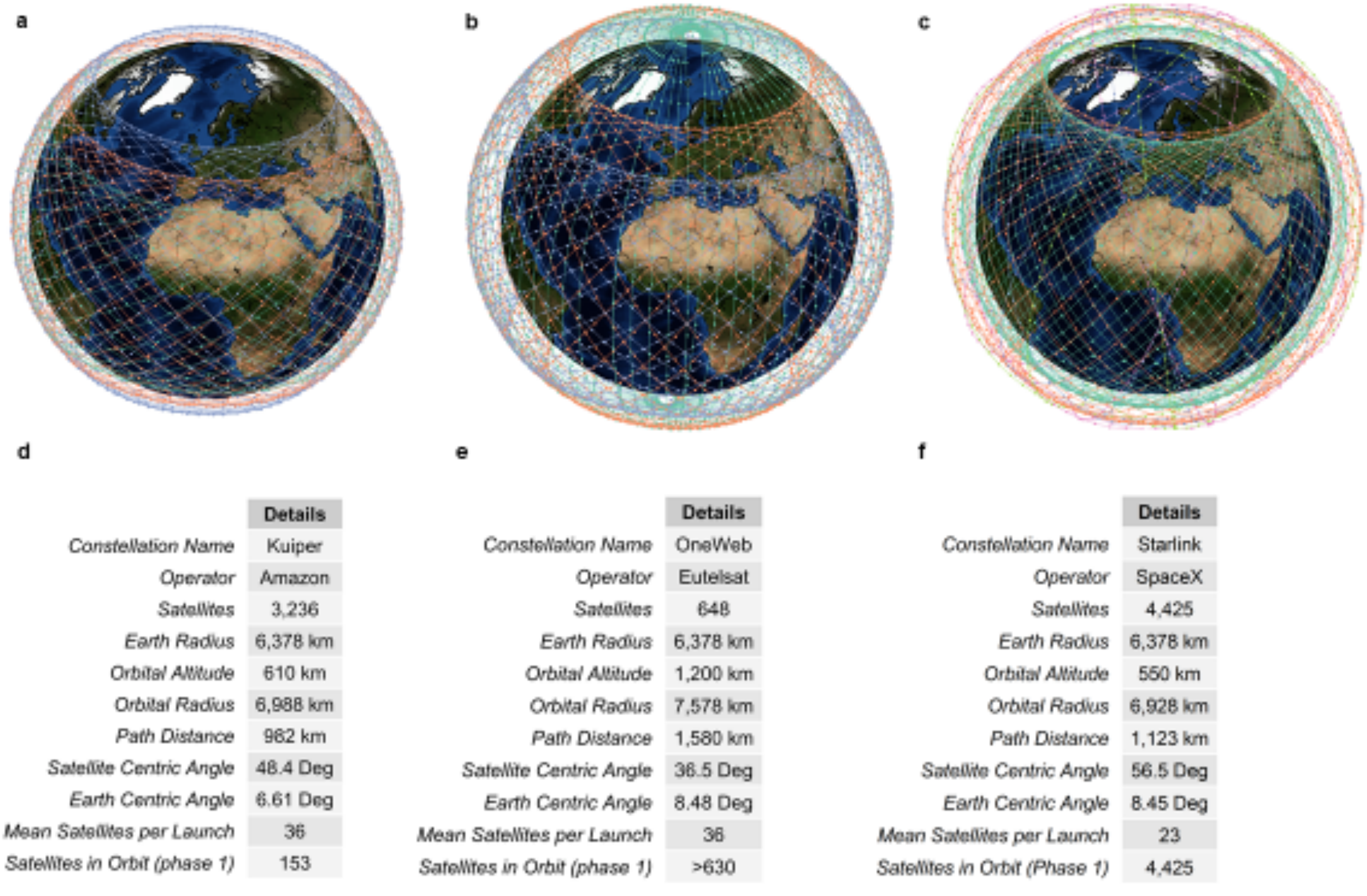

| | Details |
|---|---|
| Constellation Name | Kuiper |
| Operator | Amazon |
| Satellites | 3,236 |
| Earth Radius | 6,378 km |
| Orbital Altitude | 610 km |
| Orbital Radius | 6,988 km |
| Path Distance | 962 km |
| Satellite Centric Angle | 48.4 Deg |
| Earth Centric Angle | 6.61 Deg |
| Mean Satellites per Launch | 36 |
| Satellites in Orbit (phase 1) | 153 |

| | Details |
|---|---|
| Constellation Name | OneWeb |
| Operator | Eutelsat |
| Satellites | 648 |
| Earth Radius | 6,378 km |
| Orbital Altitude | 1,200 km |
| Orbital Radius | 7,578 km |
| Path Distance | 1,580 km |
| Satellite Centric Angle | 36.5 Deg |
| Earth Centric Angle | 8.48 Deg |
| Mean Satellites per Launch | 36 |
| Satellites in Orbit (phase 1) | >630 |

| | Details |
|---|---|
| Constellation Name | Starlink |
| Operator | SpaceX |
| Satellites | 4,425 |
| Earth Radius | 6,378 km |
| Orbital Altitude | 550 km |
| Orbital Radius | 6,928 km |
| Path Distance | 1,123 km |
| Satellite Centric Angle | 56.5 Deg |
| Earth Centric Angle | 8.45 Deg |
| Mean Satellites per Launch | 23 |
| Satellites in Orbit (Phase 1) | 4,425 |

**Fig. 1 | Technical details of the LEO constellations (as of December, 2023). a,** Amazon's Kuiper has launched 153 satellites (5%) ot its planned 3,236 phase 1 satellites, **b,** OneWeb has deployed 98% of the 648 phase 1 satellites, **c,** SpaceX Starlink has launched 100% of the 4,425 phase 1 satellites (Federal Communication Commission, 2021; Federal Communications Commission, 2020b, 2017).

### *2.1. Life Cycle Assessment of satellite constellations*

Emissions produced during the launching of satellites depend on the rocket vehicle used. Most operators planning or launching LEO broadband satellites have used (or intend to use) SpaceX's Falcon-9 or Falcon-Heavy, the European Space Agency's (ESA's) Ariane, or Russia's Soyuz-FG rocket launch systems. This analysis focuses predominantly on the emissions produced by rocket and propellant manufacturing, transportation, rocket testing and finally launching carried satellite payloads into LEO (and compared to a hypothetical GEO system).

Fig. 2 illustrates these rocket vehicles along with key technical specifications. Depending on the properties of each launch capability, either one or a combination of propellants is utilized, leading to a unique profile of environmental emissions when ignited. Importantly, many compounds are released into the environment, including nitrogen gas, carbon dioxide, carbon monoxide, black carbon, water vapor, hydrogen gas, aluminum oxide, hydrochloric acid and the radicals of chlorate, hydrate and nitrate (Ross et al., 2010; Wilson, 2019).

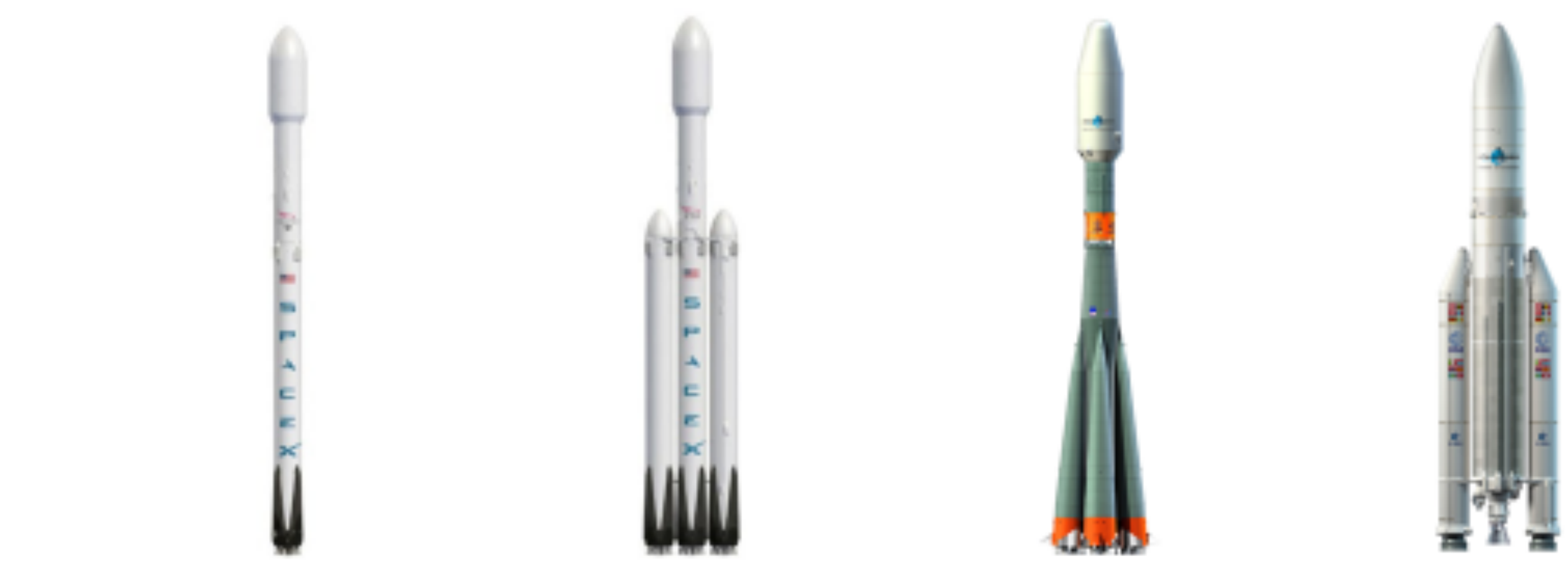

| | Falcon-9 | Falcon-Heavy | Soyuz-FG | Ariane-5 |
|---|---|---|---|---|
| *Dry Mass* | 22 tonnes | 66 tonnes | 14 tonnes | 53 tonnes |
| *Operator* | SpaceX | SpaceX | Progress Rocket Space Center | European Space Agency |
| *Payload to LEO* | 23 tonnes | 64 tonnes | 8 tonnes | 20 tonnes |
| *Fuel Type* | Kerosene (Rocket Propellant-1) | Kerosene (Rocket Propellant-1) | Kerosene and Hypergolic | Solid, Hypergolic and Cryogenic |
| *Height* | 70 m | 70 m | 46 m | 46 - 52 m |
| *Price Per Launch* | US$ 62 Million | US$ 90 Million | US$ 80 Million | US$ 149-198 Million |
| *Propellant Mass* | 488 tonnes | 1,397 tonnes | 256 tonnes | 7,645 tonnes |
| *Fuel* | Hydrocarbon | Hydrocarbon | Hydrocarbon | Hydrogen |

**Fig. 2 | Details of the rocket launching vehicles used by LEO constellations.** None of the constellations have hitherto used the Falcon-Heavy rocket. Starlink Phase 2 will likely be launched via Starship, which is still being developed and has substantially different technical details. OneWeb ceased using the Russian Soyuz-FG for launching satellite payloads in 2022[8,9,10].

Quantifying these emissions from rocket launches is complex and not well understood. However, there has been extensive work conducted on approximating the emissions per mass of the fuel burned for the four common propellants used(Ross et al., 2010, 2009; Ross and Sheaffer, 2014) and the damage of these compound emissions have on the environment (Miraux et al., 2022; Schulz and Glassmeier, 2022; Barker et al., 2024; Brown et al., 2024; Dominguez Calabuig et al., 2024).

Given the differences in constellation size, rocket launch vehicles, quantity of rocket launches, and the provided broadband capacity, each constellation has heterogenous environmental impacts, as reported here. The impacts are broken down in five categories of different LCA units. Firstly, Global Warming Potential (GWP) defined as the radiative forcing in carbon dioxide equivalents ($CO_2$eq.) over a 100-year horizon by the Intergovernmental Panel on Climate Change . Secondly, Ozone Depletion Potential (ODP) defined by the World Meteorological Organization (WMO) (World Meteorological Organization (WMO), 2022) as the steady-state depletion potential in chlorofluorocarbon-11 equivalents (CFC-11eq). Thirdly, Mineral & Metal Resource Depletion Potential defined as the abiotic resource depletion (reserve base) in antimony (Sb) equivalents as implemented by the Centrum voor Milieuwetenschappen (CML) at the University of Leiden (Guinée et al., 2022a; van Oers et al., 2002; World Meteorological Organization (WMO), 2022), and recommended by the ESA Life Cycle Assessment (LCA) Handbook (Guinée et al., 2022b).

[8] ArianneSpace Group, 2022. Ariane 5: The Heavy Launcher. URL https://www.arianespace.com/vehicle/ariane-5/ (accessed 9.14.22).
[9] Zak, A., 2022. Soyuz-FG's long road to retirement. URL https://www.russianspaceweb.com/soyuz-fg.html (accessed 9.14.22).
[10] SpaceX LLC, 2025. Falcon User's Guide. URL https://www.spacex.com/assets/media/falcon-users-guide-2025-05-09.pdf (accessed 11.6.2025)

Fourthly, Freshwater Aquatic Ecotoxicity Potential as the Comparative Toxic Units for ecosystems (CTUe) as implemented in USEtox in potentially affected fraction of species per $m^3$ per day ($PAF.m^3.day$). Finally, Human Toxicity Potential as the Comparative Toxic Units for humans (CTUh), as implemented in USEtox as the estimated increase in morbidity (Cases).

## 3. Methodology

To assess the sustainability implications of different LEO megaconstellations, we developed the open-source Sustainability Analytics for Low Earth Orbit Satellites (Saleos) codebase. In this method, we describe each step in the Saleos modeling process taken to estimate the incurred environmental sustainability impacts, provided capacity, potential demand, and associated financial and social costs, applied here to the three main LEO constellations (as well as a comparable GEO constellation). This integrated modeling approach is detailed further in the Supplementary Information, demonstrating how each of these steps fit together, given the salient exogenous and endogenous model variables used to produce the results (see Fig. 3).

### *3.1. Life Cycle Assessment*

A process-based LCA is utilized to quantify the environmental impacts associated with delivering the necessary satellites to complete each phase 1 LEO constellation. LCA is a technique used to model the environmental impacts of a process, product, or service over their entire life, from raw material extraction through to the end of lifetime of each asset (internationally standardized via ISO 14040[11] and ISO 14044[12]. The process-based approach is centered on scientifically analyzing specific activities (i.e., mass/material balance, scientific characteristics, etc.) and linking these to a functional unit. A functional unit describes the quantity of a product or product system based on the performance it delivers in its end-use application. In this case, the functional unit refers to the total number of launches required to place all proposed satellites within each constellation into their desired orbit. The activities accounted for under this functional unit are determined based on the system boundary in the Supplementary Information.

The data on production of different rockets used for launching satellites are sourced from the Strathclyde Space Systems Database (SSSD). The SSSD has a variety of datasets on the production of different launchers, including Falcon-9, Ariane, and Soyuz-FG, carefully formed based on freely available industry data and interviews with a variety of relevant industrial stakeholders. We use generic hydrogen (HYD) and hydrocarbon (HYC) rocket vehicles when the exact launcher is not in the SSSD database or the rocket type is unknown. These generic rockets use mean values produced from fully modeled launchers within the SSSD. In total for this evaluation, 63% of assessed LEO launches are based on a modeled rocket, 36% utilize a generic rocket for a known launcher, and only 2% utilize a generic rocket for an unknown launcher. A similar approach is utilized for GEO. The Supplementary Information specifies the full LCA method.

[11] International Organization for Standardization, 2023. ISO 14040:2006: Environmental management — Life cycle assessment — Principles and framework. URL https://www.iso.org/obp/ui/#iso:std:iso:14040:ed-2:v1:en (accessed 6.7.23).

[12] International Organization for Standardization, 2023. Environmental management — Life cycle assessment — Requirements and guidelines, 2014. Environmental management — Life cycle assessment — Requirements and guidelines. URL https://www.iso.org/standard/38498.html (accessed 6.7.23).

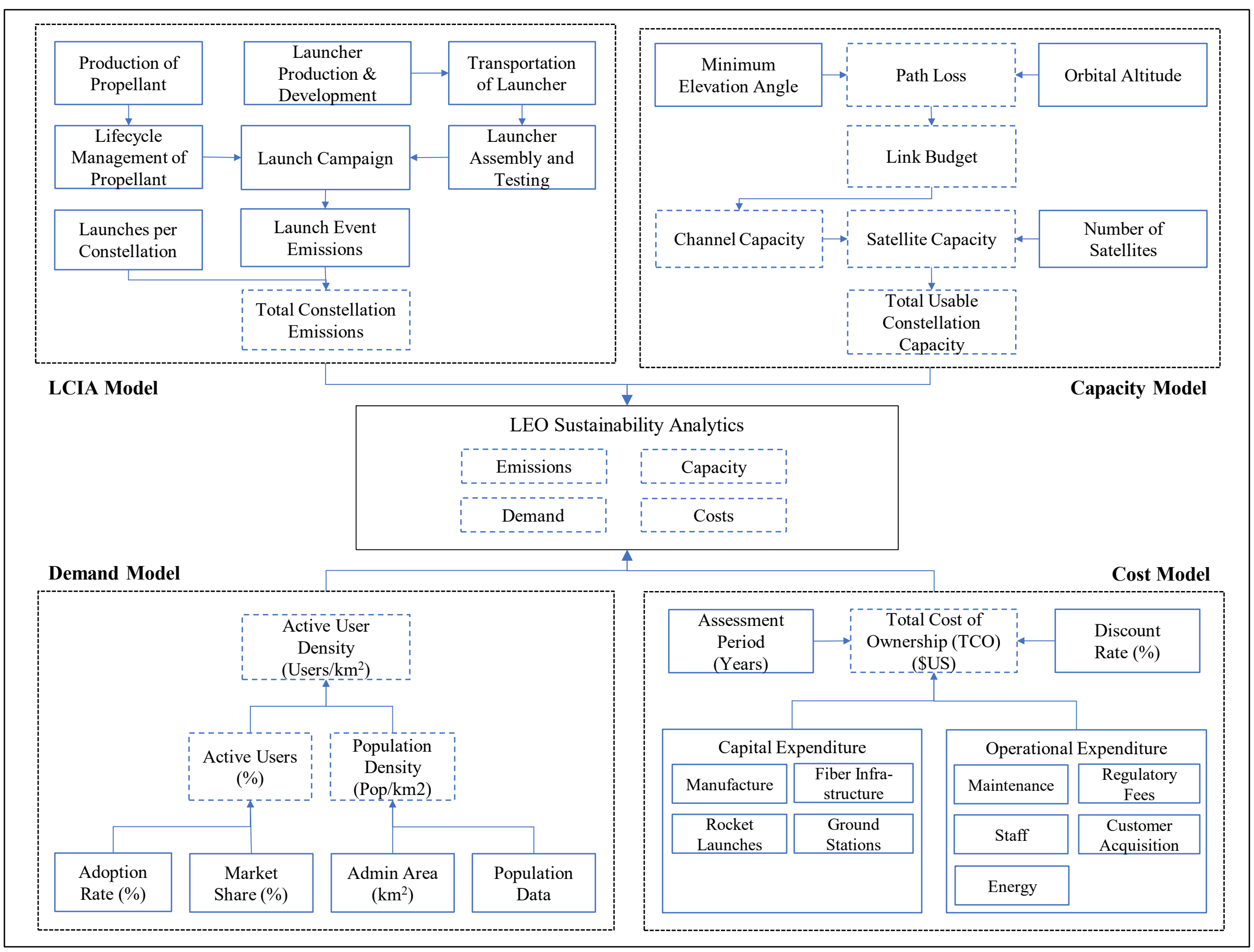

**Fig. 3 | Method box diagram.** Outlines environmental, capacity, demand and financial cost modules.

## *3.2. Provided capacity*

The downlink channel capacity of each LEO constellation is estimated as this is generally the main bottleneck for subscribers trying to access Internet content. To do this, the Friss Transmission equation is utilized, as detailed in the Supplementary Information, following an established methodology (Biscarini et al., 2022; Giggenbach et al., 2023; Gongora-Torres et al., 2022; Muttiah, 2023; Popescu, 2017; Saeed et al., 2021). Firstly, information is gathered on antenna characteristics and then used to estimate the energy per bit to noise power spectral density ratio $\left(\frac{E_b}{N_O}\right)_{dB}$. Based on FCC filings, the channel capacity is calculated from the modulation coding schemes and spectral efficiency values, with the expectation that next generation satellites are likely to use Adaptive Coding and Modulation (ACM). Next, the total satellite capacity (in Mbps) is obtained, by multiplying the channel capacity by the number of beams and channels. Finally, the total usable constellation capacity is estimated by multiplying the total satellite capacity, by the number of satellites in each constellation, along with a factor which represents the average time each satellite spends over land serving subscribers (as opposed to generally idle over ocean).

### 3.3. *Potential demand*

We develop scenarios of future change which capture demand uncertainty in adoption. Estimating future demand is a key challenge when evaluating sustainability aspects of infrastructure systems (Hall et al., 2016), raising the need for scenarios, given the lack of available robust scientific information for modeling. Information is gathered on the current number of LEO broadband subscribers by Q4 2023, and is generally used as the low adoption scenario. As detailed thoroughly in the Supplementary Information, the baseline and high adoption scenarios see the existing customer base broadly increase by 1.5x and 2x, respectively, following industry information. The estimated data rate capacity per subscriber (Mbps) can then be obtained, by dividing the total usable capacity with the number of subscribers in each scenario. Finally, the maximum quantity of data traffic, which this capacity can enable each subscriber to download per month, is estimated (in GB/Month).

### 3.4. *Costs*

Both the financial and social costs associated with launching phase 1 of each LEO constellation are estimated following (Osoro and Oughton, 2021) and (Li et al., 2023), as detailed in the Supplementary Information. Firstly, the lifetime total capital expenditure (capex) is evaluated when considering the costs of satellite manufacturing, satellite launch, ground station investment, and fiber infrastructure. Next, the lifetime total operational expenditure is evaluated when considering the recurring costs of ground station energy consumption, staff labor, regulatory fees, subscriber acquisition, and maintenance. Finally, the TCO is obtained by summing all initial capital expenditure costs, and recurring annual operational expenditure, discounted at a rate of 7% based on the Cost of Capital[13] over the lifetime of each LEO/GEO constellation. To normalize for different lifetimes, metrics are converted to either annual or monthly quantities. Secondly, the total climate change emissions (the GWP) associated with each LEO constellation is multiplied by the social cost of a single tonne of carbon as established in (Rennert et al., 2022a), to obtain the SCC.

## 4. Results

Different rocket combinations have been or will be used to launch upcoming satellite constellations, as detailed in Fig. 4a. Currently, Starlink has made 127 launches to place all its 4,425 satellites in orbit using Falcon-9 a hydrocarbon (HYC) fuel-based rocket. Similarly, OneWeb placed 96 of its satellites in 3 launches with Falcon-9, while the remaining were made on India’s LVM3 (2 launches for 72 satellites) (HYD) and Russian Soyuz-FG (14 launches for 394 satellites) (HYC). For Kuiper, Amazon has announced the majority of their future launches, including 38 via United Launch Alliance’s Vulcan Centaur rocket (HYC), 18 via Arianespace’s Ariane-6 hydrogen (HYD) rocket, up to 27 via Blue Origin’s New Glenn (HYC) rocket, and 3 via Falcon-9[14]. We split the remaining 4 between generic hydrocarbon (HYC) and generic hydrogen (HYD) rockets. Finally, for the hypothetical GEO operator, we model 10 launches via HYC and 9 launches via HYD (with one satellite per launch).

The resulting annual emissions per subscriber (in kg $CO_2$eq) are illustrated in Fig. 4b. Environmental impacts are commonly reported by subscriber for telecommunication networks annually (Lundén et al., 2022a;

[13] New York State University, 2023. Cost of Capital. URL https://pages.stern.nyu.edu/~adamodar/New_Home_Page/datafile/wacc.html (accessed 6.30.23).

[14] Amazon, 2023. Amazon secures 3 launches with SpaceX to support Project Kuiper deployment. US Amazon. URL https://www.aboutamazon.com/news/innovation-at-amazon/amazon-project-kuiper-spacex-launch (accessed 2.28.24).

Malmodin et al., 2024; Ruiz et al., 2022). This is an essential way to provide decision makers with an understanding of system impacts, while accounting for (i) the quantity of users receiving service, and (ii) different asset lifetimes (Kelly and Comendant, 2021). Publicly available subscriber data (Jewett, 2022; Sheetz, 2022) are utilized to account for the low, baseline and high adoption scenarios, as detailed in the methodology and Supplementary Information (SI). This baseline includes 2.5 million future subscribers for Kuiper (0 currently), 0.8 million for OneWeb (about 0.2 million currently), 3.5 million for Starlink (2.2 million as at Q4 of 2023), and 2.5 million for the representative GEO operator. Estimated annual emissions correspond to 303±131 kg $CO_2$eq/subscriber for Kuiper, 274±101 kg $CO_2$eq/subscriber for OneWeb, 172±51 kg $CO_2$eq/subscriber for Starlink, and 21±9 kg $CO_2$eq/subscriber for GEO. Thus, on average the subscriber emissions from LEO constellations are more than 12 times higher than the representative GEO operator.

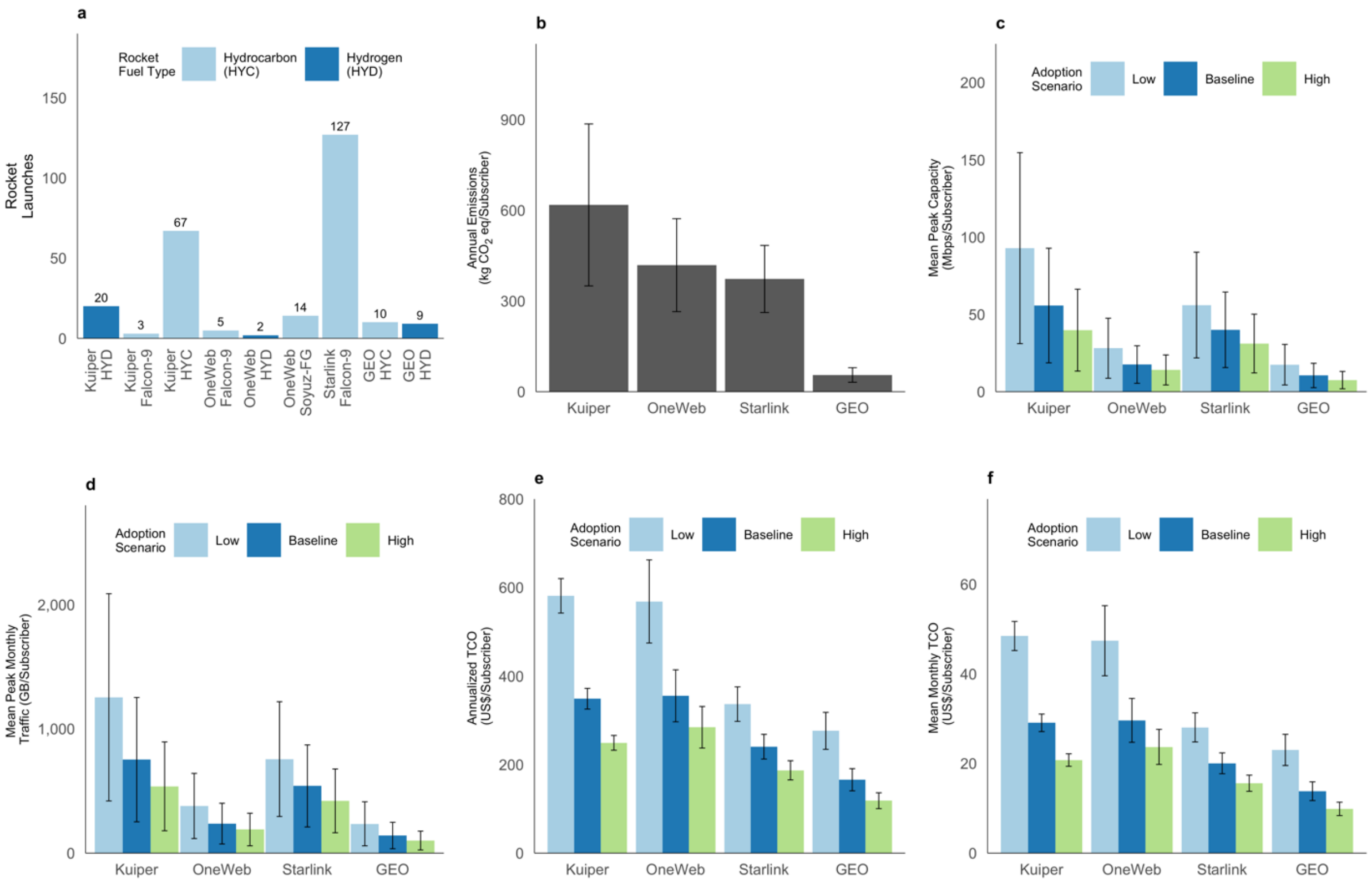

**Fig. 4 | Constellation Metrics. a,** Quantity of rocket launches by constellation and rocket fuel-type, based on information as of December 2023. We use generic hydrogen (HYD) and hydrocarbon (HYC) rocket vehicles when the exact launcher is not in the SSSD database or the rocket type is unknown (via a 50-50 split). In total for this evaluation, 62.6% of assessed LEO launches are based on a modeled rocket, 35.7% utilize a generic rocket for a known launcher, and only 1.7% utilize a generic rocket for an unknown launcher. **b,** Equivalent annual emissions estimated per subscriber for each constellation, with Confidence Intervals (CIs) representing low and high subscriber adoption scenarios, **c,** The estimated mean provided peak data rate with CIs representing 1 Standard Deviation (SD) in mean capacity for each adoption scenario, **d,** Potential peak monthly

traffic per subscriber estimated with CIs representing 1 SD in monthly traffic for each adoption scenario, **e,** The estimated annualized Total Cost of Ownership (TCO) per subscriber by constellation with CIs representing 1 SD in TCO for each adoption scenario, **f,** The estimated average monthly TCO per subscriber with CIs representing 1 SD for each adoption scenario.

Importantly, LEO constellations are aiming to serve hard-to-reach locations in rural and remote areas, where terrestrial broadband infrastructure deployment is unviable. Evaluation of operational carbon emissions suggests annual emissions intensities of 32.8 kg $CO_2$eq/subscriber in rural areas, and 39.5 kg $CO_2$eq/subscriber in remote areas (Ruiz et al., 2022) for terrestrial mobile broadband (4G). Therefore, furthering the results visualized in Fig. 4b, this means that compared to terrestrial mobile broadband, LEO is approximately 8 times higher per rural subscriber, or 6 times higher per remote subscriber. In contrast, GEO emissions of 22 kg $CO_2$eq/subscriber are nearly one third lower for rural subscribers, and nearly 50% lower for remote subscribers, for terrestrial mobile broadband (4G).

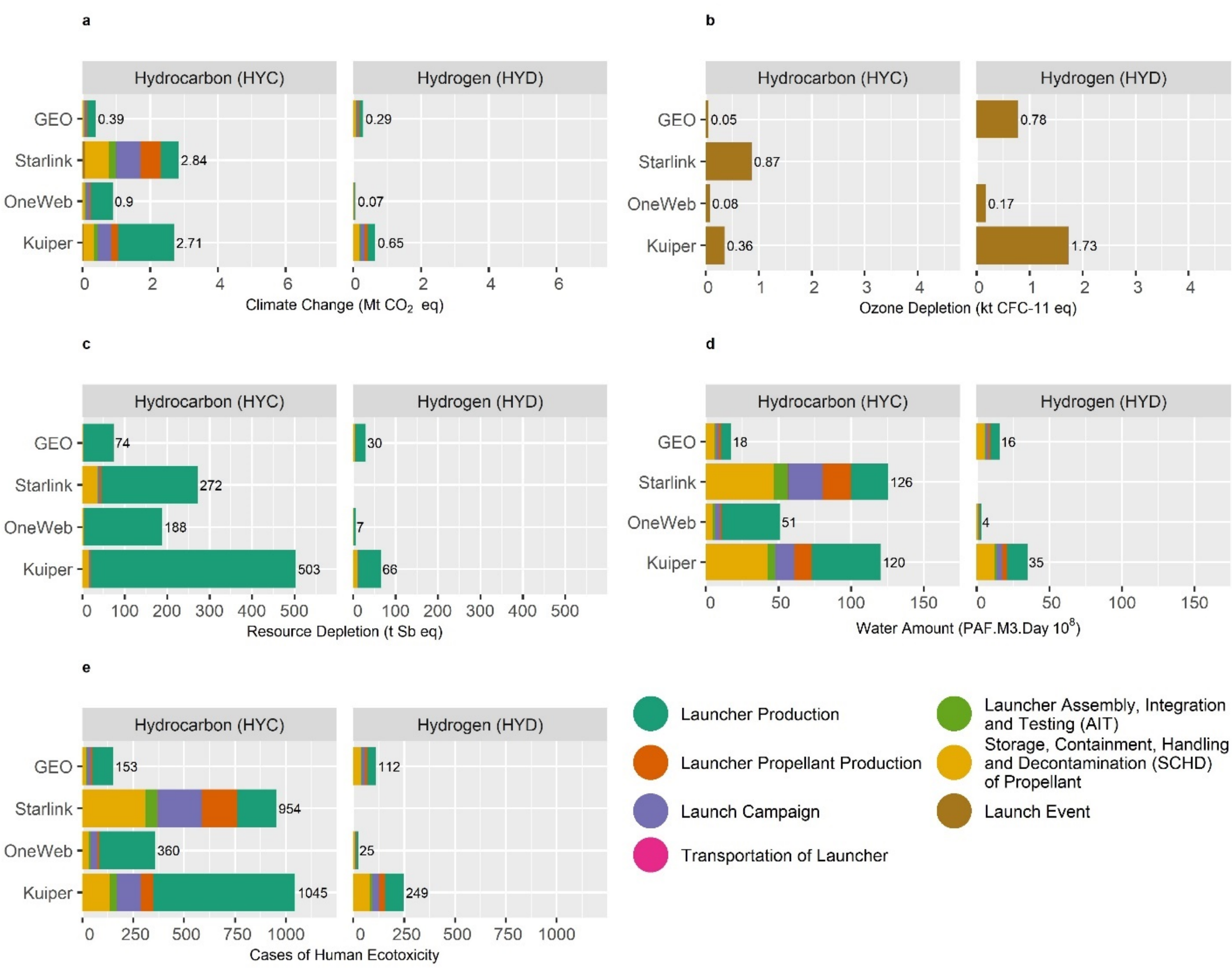

**Fig. 5 | Key constellations by environmental impact category. a**, Climate change impacts, **b,** Ozone depletion, **c,** Resource depletion, **d,** Freshwater ecotoxicity, **e,** Human toxicity.

Results are reported in Fig. 5, broken down by HYC and HYD rockets over each constellation lifetime. Carbon emissions are visualized in Fig. 5a. The results suggest that climate change emissions account for one of the highest LCA impacts. For example, for HYC the full launch of the planned Kuiper constellation is estimated to produce 2.71 Mt $CO_2$eq, versus 0.9 Mt $CO_2$eq for OneWeb, 2.84 Mt $CO_2$eq for Starlink and 0.39 Mt $CO_2$eq for GEO as illustrated in Fig. 5a. Whereas, the associated emissions for the HYD portion are comparatively lower for Kuiper (0.65 Mt $CO_2$eq), OneWeb (0.07 Mt $CO_2$eq) and GEO (0.29 Mt $CO_2$eq). For comparison, terrestrial European mobile operators (2G-4G) reported annual emissions of approximately 3.4 Mt in 2018 (Lundén et al., 2022b) for 401 million subscriptions (highlighting the need for per subscriber metrics, as presented in Fig. 4b).

In accounting for ozone depletion, when considering a HYC rocket, 0.36 Kt, 0.08 Kt, 0.87 Kt and 0.05 kt of CFC-11eq are estimated to be produced when all satellites are launched by Kuiper, OneWeb, Starlink and the GEO operator, respectively (Fig. 5b). In using a HYD rocket, the emissions are 1.73 Kt, 0.17 Kt and 0.78 Kt for Kuiper, OneWeb and GEO, respectively. Planetary boundaries define environmental limits within which humanity can safely operate to maintain a sustainable human presence on Earth. They essentially act as the ecological threshold based on the level of impact that can be sustained in a single year, which is important because the future of the space sector will be constrained by environmental limits (Miraux, 2022b). When comparing these CFC-11eq values against annual planetary boundaries, the value provided for Starlink represents 0.002% of the annual planetary boundary for ozone depletion (Sala et al., 2017), whilst Kuiper, OneWeb and GEO contribute just 0.0004%, 0.0001% and 0.0002%. See the Supplementary Information LCA results section for further review of all metrics (resource depletion, freshwater and human toxicity).

Lastly, the SCC measures the monetary value of damages to society caused by emitting an incremental ton of CO2 or its equivalents over this unit's lifetime in the atmosphere (Rennert et al., 2022b). This approach is used in conducting cost-benefit analysis of policies which may have sustainability impacts (often required by regulatory agencies) (U. S. Government Accountability Office, 2022). Monetization via SCC enables assessment of sustainability and economic impacts in common units and does not represent legal claims. Here, we estimate the SCC for the two emissions scenarios associated with phase 1 of each constellation, as illustrated in Fig. 6.

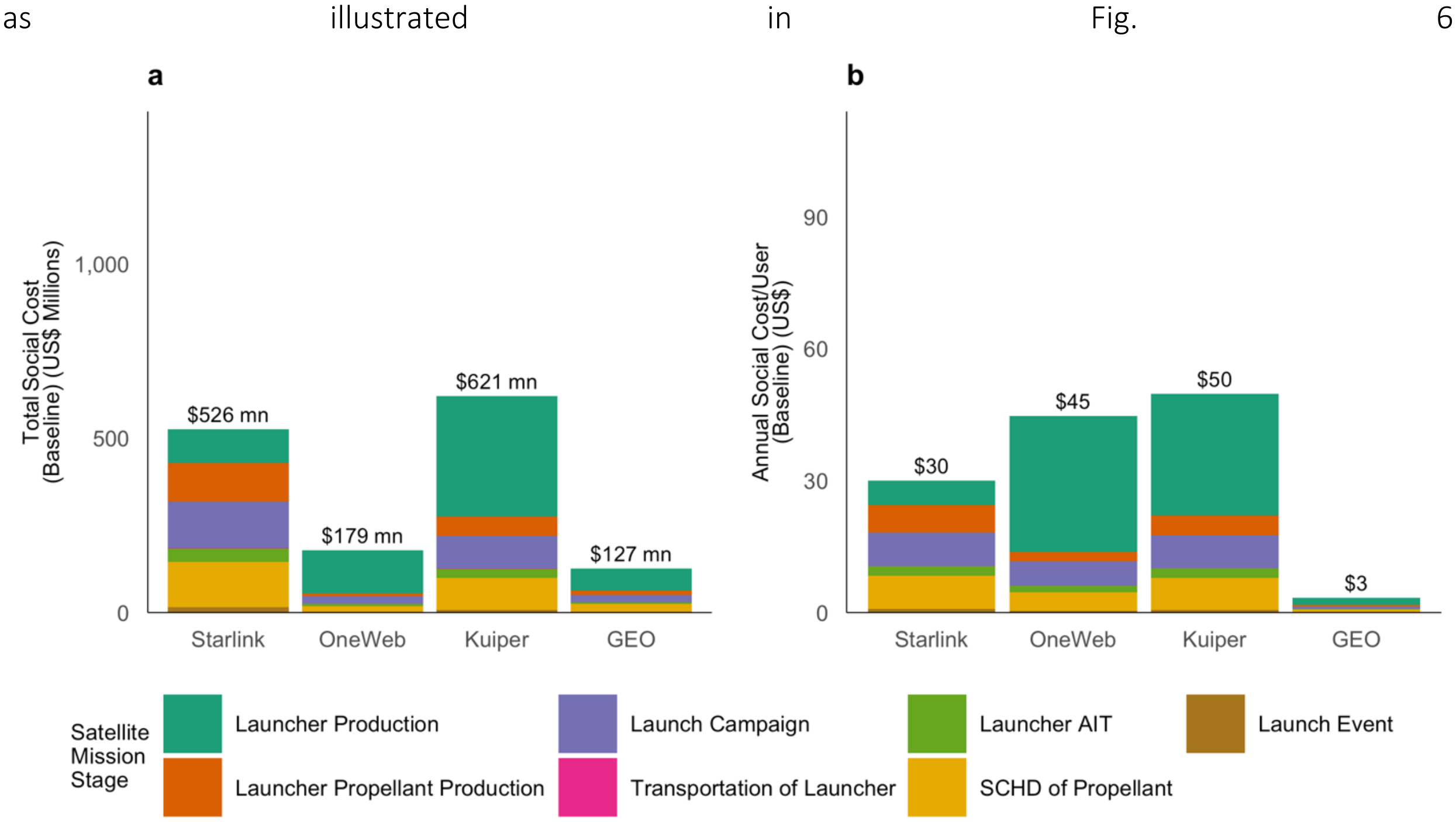

**Fig. 6 | Social cost of carbon. a**, The total SCC of the emissions over a five-year (LEO) and fifteen-year (GEO) time horizon, **b,** Annualized SCC per subscriber.

Lastly, the SCC measures the monetary value of damages to society caused by emitting an incremental ton of $CO_2$ or its equivalents over this unit's lifetime in the atmosphere (Rennert et al., 2022b). This approach is used in conducting cost-benefit analysis of policies which may have sustainability impacts (often required by regulatory agencies) (U. S. Government Accountability Office, 2022). Monetization via SCC enables assessment of sustainability and economic impacts in common units and does not represent legal claims. Here, we estimate the SCC for the two emissions scenarios associated with phase 1 of each constellation, as illustrated in Fig. 6. Firstly, the total social cost is estimated at $621 million for Kuiper, versus $179 million for OneWeb, $526 million for Starlink and $127 million for a representative GEO operator (Fig. 6aLastly, the SCC measures the monetary value of damages to society caused by emitting an incremental ton of CO2 or its equivalents over this unit's lifetime in the atmosphere (Rennert et al., 2022b). This approach is used in conducting cost-benefit analysis of policies which may have sustainability impacts (often required by regulatory agencies) (U. S. Government Accountability Office, 2022). Monetization via SCC enables assessment of sustainability and economic impacts in common units and does not represent legal claims. Here, we estimate the SCC for the two emissions scenarios associated with phase 1 of each constellation, as illustrated in Fig. 6. ). Secondly, it is imperative that these estimates are broken down by the number of subscribers expected to be served by each constellation annually. For example, the per subscriber social cost is estimated to be $50 for Kuiper in the baseline adoption scenarios, versus $45 for OneWeb, $30 for Starlink and $3 for GEO operator (Fig. 6bLastly, the SCC measures the monetary value of damages to society caused by emitting an incremental ton of CO2 or its equivalents over this unit's lifetime in the atmosphere (Rennert et al., 2022b). This approach is used in conducting cost-benefit analysis of policies which may have sustainability impacts (often required by regulatory agencies) (U. S. Government

Accountability Office, 2022). Monetization via SCC enables assessment of sustainability and economic impacts in common units and does not represent legal claims. Here, we estimate the SCC for the two emissions scenarios associated with phase 1 of each constellation, as illustrated in Fig. 6. ).

Environmental policies utilizing these estimates will likely impose costs on the space industry. Some of these costs will be passed on to service subscribers, reducing service availability for those who need broadband, and would benefit from progressing the SDGs. These carbon prices offer useful guidance on the magnitude of the externalities these systems generate. Balancing the management of these externalities against the social benefits of greater broadband access is a challenging task requiring further development of integrated modeling frameworks as presented here.

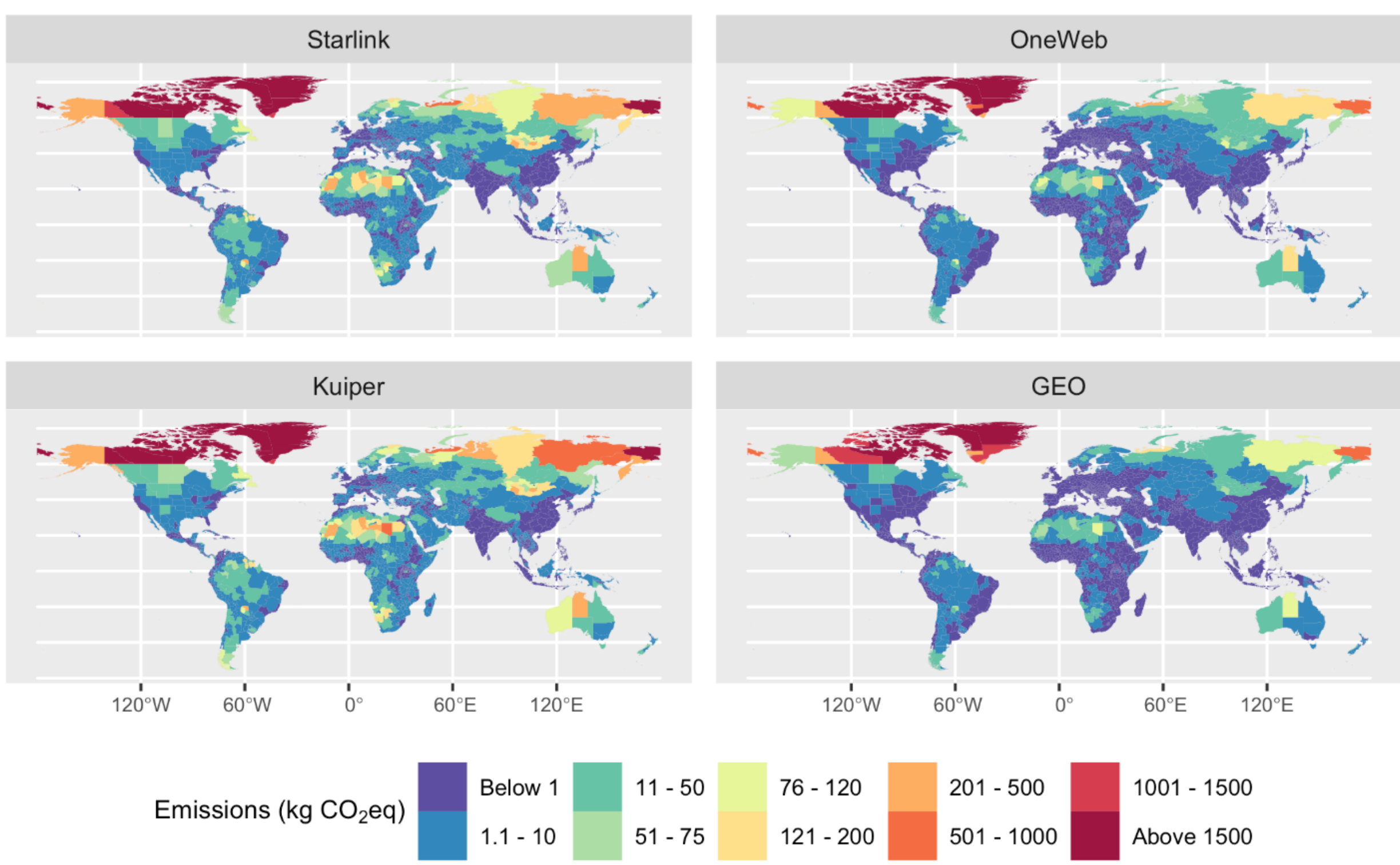

**Fig. 7 | Emissions per user.** Averaged emissions per user for different sub-regional areas considering the baseline adoption scenario across all the constellations.

In Fig. 7, we present the emissions per user for different sub-regional areas across the globe. First, constellation with large number of satellites require more launches leading to higher emissions (e.g Starlink 2.84 Mt $CO_2$eq compared to OneWeb's 0.9 Mt $CO_2$eq). When these total emissions are disaggregated across users in an area, ultra-dense constellations such as Starlink and Kuiper results in large per user emissions across the coverage areas. It is important to note that the emissions per user is higher in sparsely populated areas such as deserts, around rainforests and the poles. This is well demonstrated with emissions above 120 kg $CO_2$eq in remote regions of Canada, Russia, Amazon rainforests, Australian, Kalahari and Sahara deserts. Equally, the emissions per user falls below 1 kg $CO_2$eq in densely populated areas such as large parts of Asia (India and China), Western Europe, Central America (e.g. Mexico) and Sub-Saharan Africa.

As opposed to LEO constellations, the overall emissions per user considering a GEO system is generally low across the globe (below 10 kg $CO_2$eq). This explains the limited research interest in the environmental sustainability of providing satellite broadband through GEO systems. For instance, Fig. 7, illustrates that the average emissions per user across Africa is below 50 kg $CO_2$eq for the case of GEO. However, when considering a LEO system such as Starlink to provide coverage, some areas record an average emissions per user of above 500 kg $CO_2$eq.

## 5. Analysis and Discussion

The results presented demonstrate that the phase 1 LEO constellations currently being deployed have significant sustainability implications, with these impacts likely to substantially increase as the sector aims to move to constellations an order-of-magnitude larger over the next decade (from thousands of satellites to tens of thousands). Hitherto, the space sector has largely operated and been regulated under the premise that launch traffic and operational intensity is low enough to minimize environmental impacts (Gilbert and Vidaurri, 2020; World Meteorological Organization (WMO), 2022). Our analysis shows this assumption is breaking down in the era of megaconstellations, given the thousands of planned satellite assets requiring frequent rocket launches to reach orbit.

Indeed, phase 1 LEO constellations have annual operational environmental footprints (0.2-0.7 Mt $CO_2$eq) equivalent to the energy usage of 24-85k annual US homes or 43-150k annual gasoline-powered passenger vehicles. The annual subscriber environmental footprints for LEO (172-303 kg $CO_2$eq) are equivalent to a one-way economy-class flight between London and Milan (257 kg $CO_2$eq) (900 km). While improvements in launcher designs and transportation logistics may reduce this footprint, the coming rush of large LEO constellations suggests the total environmental footprint of the space sector is likely to rise regardless (particularly as this assessment did not include a range of other growing space activities, such as tourism). The comparative GEO constellation had relatively modest annual emissions impacts ranging from 21-55 kg $CO_2$eq/subscriber, similar to driving from Florence to Bologna (117 km).

Much of the focus on LEO megaconstellation hitherto has been regarding orbital debris and changes to the night sky (Rao et al., 2020). Those impacts are not generally covered under existing environmental policies and international agreements. By contrast, the environmental sustainability impacts we measure here are not novel *per se*, so are better represented in existing environmental policy. While $CO_2$ emissions may not be covered under binding international agreements, they are recognized under existing legal structures, e.g., the Paris Agreement. Certainly, further research on LCA impacts of satellite constellations is needed to clarify their implications for existing environmental agreements and targets.

How might these responsibilities be carried out? Broadly, there are two paths: Targeting launches within a covered jurisdiction directly, or targeting services provided to subscribers. While some concerns may exist regarding polluting launchers fleeing to jurisdictions with laxer regulations (i.e. “launch leakage”), some prior analyses of environmental regulations have found relatively low levels of leakage (Levinson and Taylor, 2008). The magnitude of this effect in the space sector is an important open empirical question. Where targeting launchers is infeasible or not currently taking place, existing border carbon adjustment policies[15] offer an example of a potential policy response (Copley, 2023; Fischer and Fox, 2012; Fontagné and Schubert, 2023). By pricing emissions at the point of service delivery, national actors – particularly those with large or lucrative domestic markets – can partially offset the incentive to flee to so-called "pollution havens".

## 6. Conclusions and Limitations

This assessment finds that LEO constellations provide substantial capacity improvements in broadband Internet for rural and remote communities to access. However, this comes at a price, as emissions from LEO constellations are quite considerably higher compared to serving rural and remote communities via terrestrial 4G mobile networks, based on the plausible demand scenarios evaluated. For example, on average, launching broadband LEO constellations results in 250 kg $CO_2$eq/subscriber annually, roughly 6-8 times higher than values for terrestrial mobile networks (with comparative values of 32.8 kg $CO_2$eq per rural subscriber and 39.5 kg $CO_2$eq per remote subscriber).

Secondly, we find that compared to a representative GEO constellation, LEO constellations are approximately 9-12 times more emissions intensive. For example, GEO incurs approximately 21-55 kg $CO_2$eq/subscriber, which are within the same order-of-magnitude as serving rural and remote subscribers via terrestrial 4G mobile broadband. In contrast though, the mean peak capacity provided by LEO constellations is on average four times higher under the plausible baseline demand scenario (11 Mbps/subscriber for GEO versus 39 Mbps/subscriber for LEO, if all users simultaneously access the network).

Currently, this study only focuses on phase 1 of Amazon Kuiper, OneWeb and SpaceX Starlink, while the planned phase 2 constellations are an order-of-magnitude larger, raising the need for greater consideration of space sector environmental impacts and future research on quantifying phase 2 emissions. Indeed, space companies and regulators require comprehensive sustainability analytics, as presented here, to inform mitigation efforts capable of balancing environmental effects, provided broadband Internet capacity, and other financial and social cost considerations. It would also be beneficial for more research to (i) reduce and quantify uncertainty in emissions estimates, especially for ozone, and (ii) to help estimate emissions impacts from re-entry particles. Future policy decision must balance the sustainability impacts of LEO constellations against the benefits of improved broadband Internet for achieving the SDGs.

We identify three key methodological limitations. The first limitation relates to the uncertainties present within the environmental modeling emissions factors, as there is still considerable scientific research to be undertaken to better understand and quantify emissions. Should new emissions factors emerge, the open-source codebase could be readily utilized to re-assess the implications with regards to LEO constellations.

[15] European Union Commission, 2023. Carbon Border Adjustment Mechanism. URL https://taxation-customs.ec.europa.eu/carbon-border-adjustment-mechanism_en (accessed 6.29.23).

The second methodological limitation relates to the wider estimation of the environmental impacts of LEO, as the system boundary utilized had a notable exclusion in the form of the production, development and testing of spacecraft. The reason this was excluded from the study is because no LCI data for this activity could be found for Amazon Kuiper, OneWeb or SpaceX Starlink, including either a list of components or bill of materials. In the future, it would be beneficial for this analysis to be revisited should such information later become publicly accessible. Finally, this assessment does not include potential re-entry emissions which is an emerging area of research. Therefore, future evaluations should look to explicitly include these impacts.

## Data and Code availability

All data are deposited in the associated zenodo data repository at https://zenodo.org/record/8102102. The code used in the Sustainability Analytics for Low Earth Orbit Satellites (saleos) is available at https://github.com/Bonface-Osoro/saleos. The repository model code is written in Python with visualization scripts produced in R. The repository code contains input data that users can customize or replace with their bespoke values to produce new results for similar systems.

## Acknowledgements

We would like to thank Geography and Geoinformation Sciences at George Mason University for providing start-up funding for the project. Additionally, we would like to thank Dr. Whitney Lohmeyer for providing advice on the satellite capacity model.

# Supplementary Information (SI)

## Method - Emissions model

The Life Cycle Inventory (LCI) is a vital step of LCA which involves the data collection and calculation procedures relating to quantifying the inputs and outputs of the studied system. The SSSD is one of only two process-based space LCA databases which exist globally, it contains more than 250 unique foreground space-specific sustainability datasets which each contain environmental, costing and social data, based on Ecoinvent and the European reference Life Cycle Database (ELCD) background inventories (Wilson, 2019). The database was developed mainly for use in early space mission design concepts as an engineering tool. Its purpose is to scientifically quantify impacts of space system concepts and use this information to lower adverse implications, without compromising technical aspects (Wilson and Vasile, 2023). This is achieved by converting physical activity data to a developed product tree, derived from assessing all the known inputs of a particular process and calculating the direct impacts associated with the outputs of that process. The data contained in the SSSD comprises of a mixture of both primary and secondary sources, validated at the ESA through a collaborative project in 2018 (Wilson, 2018). It has been used by stakeholders across three continents to calculate the environmental footprint of a variety of space missions. The SSSD is available on request to anyone with a valid and current Ecoinvent licence.

To evaluate the relevancy of the LCI, the Life Cycle Impact Assessment (LCIA) phase was then applied, associating LCI data with specific environmental impact categories and category indicators. The selection of impact categories used within this assessment was made based on those determined to be the five major 'hotspots' of space missions, as defined by the ESA (Serrano et al., 2022). In this case, the assessment model applied to quantify each impact category was based on the recommended LCIA method outlined in the ESA space system LCA guidelines (ESA LCA Working Group, 2016).

Data collected for this analysis focus on the system boundary defined in SI Fig. 1. All information obtained was readily available in the public domain and then coupled with existing datasets contained within the SSSD (Wilson et al., 2018), to categorize the LCA impacts presented in SI Table 1. The SSSD has a variety of datasets on the production of different launchers, including Falcon-9, Ariane-5, and Soyuz-FG, carefully formed based on freely available industry data and interviews with a variety of relevant industrial stakeholders. Next, we will describe the approach for modeling individual rocket emissions, before specifying the total rocket compositions for each constellation.

Firstly, reusability of launchers was considered for Falcon-9, with the expectation that one launcher is capable of ten launches. As a result, the production of just six launchers was considered, with refurbishment to bring each launcher back to launch grade comprising part of the Assembly, Integration and Testing (AIT) for each launch. SSSD single rocket impacts are visualized in SI Fig. 2, including for generic hydrocarbon (HYC) and generic hydrogen (HYD) rockets, based on the mean impacts of modeled launchers.

Next, for the production of propellant, 480,000 kg of ammonium ice/ammonium perchlorate/hydroxyl terminated polybutadiene (solid), 184,900 kg of liquid oxygen/liquid hydrogen (cryogenic) and 10,000 kg of nitrogen tetraoxide ($N_2O_4$)/monomethylhydrazine (hypergolic) were considered per Ariane launch. In comparison, 488,370 kg of liquid oxygen/Rocket Propellant (RP-1) (kerosene) was considered per Falcon-9 launch, whilst 218,150 kg liquid oxygen/RP-1 (kerosene) and 7,360 kg of $N_2O_4$/Unsymmetrical dimethylhydrazine (hypergolic) were considered per Soyuz-FG launch. The SSSD contained datasets on the production of all of these propellant formulations, based on averaged industry data.

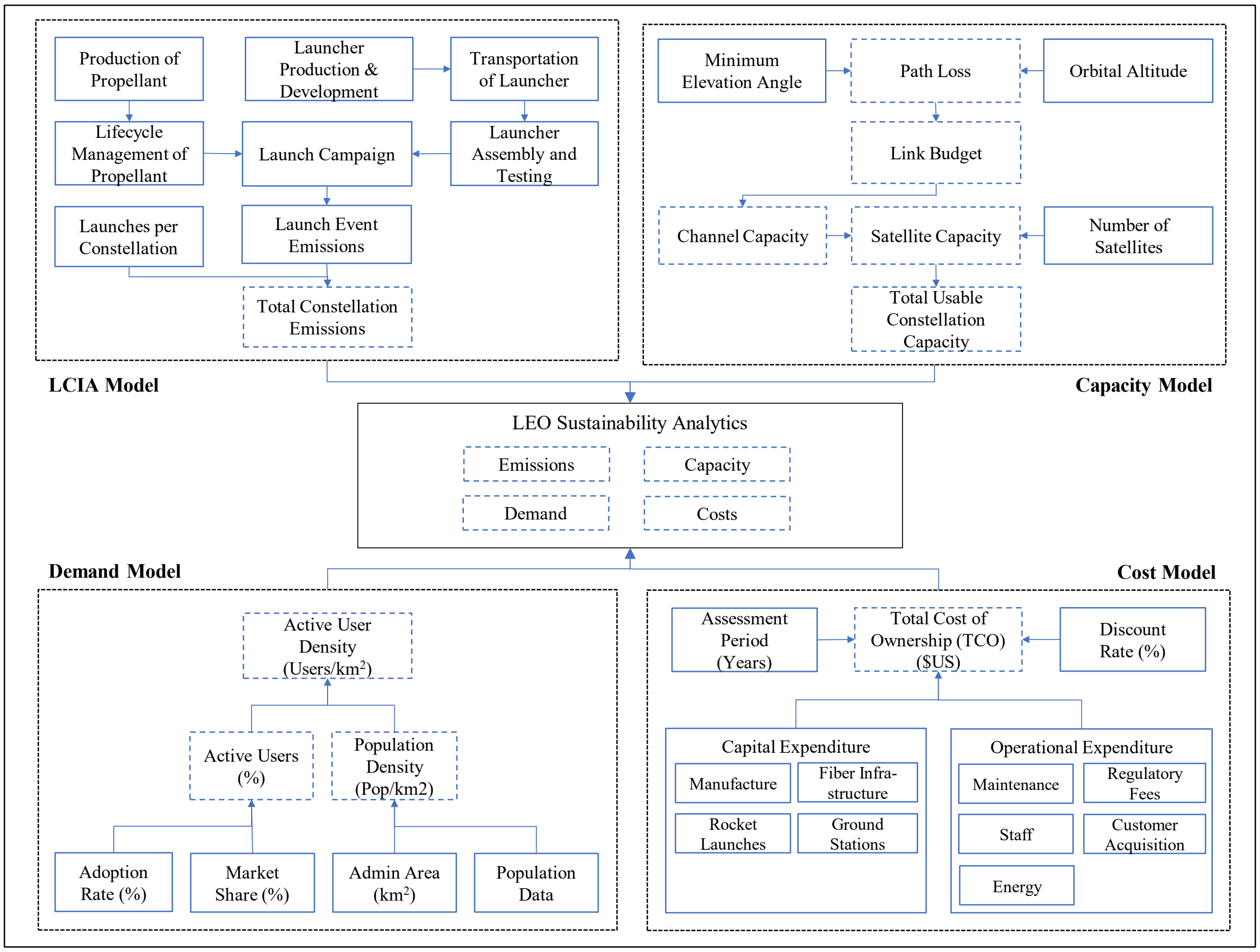

**SI Fig. 1| Method box diagram**. Outlines environmental, capacity, demand and financial cost modules.

Transportation was calculated based on the distance to bring the launcher to the launch pad. In the case of Ariane, this was the distance covered by transoceanic ship from Europe to Grand Port Maritime de la Guyane, and then via truck over land to Kourou, in French Guiana. For Falcon-9, transportation via truck was considered from the SpaceX headquarters in California to Kennedy Space Centre in Florida. The transportation of the Soyuz-FG was modeled from Samara, Russia to the launch site in Baikonur, Kazakhstan via train.

| Impact Category | Unit | Assessment Model | Method |
|---|---|---|---|
| Global Warming Potential | Kg $CO_2$eq | Bern model – Global Warming Potential over a 100-year horizon | (Myhre et al., 2013) |
| Ozone Depletion Potential | Kg CFC-11eq | WMO 1999 as implemented in the CML 2002 model | (Guinée et al., 2022) |
| Mineral & Metal Resource Depletion Potential | Kg Sbeq | CML 2002 model – Abiotic resource depletion, reserve base | (van Oers et al., 2002) |
| Freshwater Aquatic Ecotoxicity Potential | PAF.m$^3$.day | USEtox model - Comparative Toxic Units (CTUe) | (Fantke et al., 2023) |
| Human Toxicity Potential | Cases | USEtox model - Comparative Toxic Units (CTUh) | (Fantke et al., 2023) |

SI Table 1| List of the impact categories with their associated assessment methods in accordance with the ESA space system LCA guidelines.

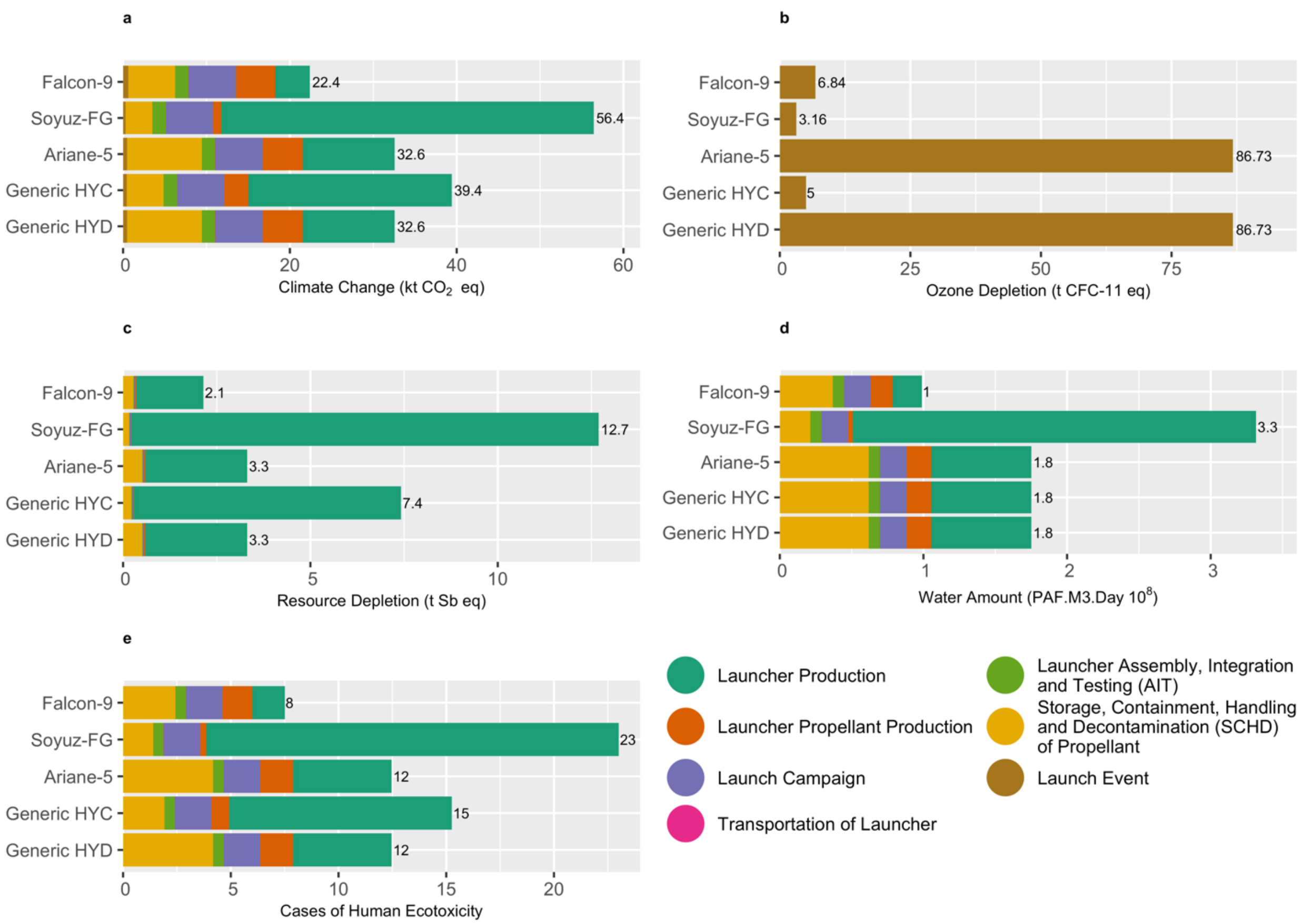

SI Fig. 2 | Single rocket LCA impacts by category. **a**, Climate change impacts, **b,** Ozone depletion, **c,** Resource depletion, **d,** Freshwater ecotoxicity, **e,** Human toxicity.

Data for the AIT of each specific launcher, along with the management of the Storage, Containment, Handling and Decontamination (SCHD) of propellant (considered for the mass/volume of all launcher propellants for a launch campaign period of 21 days), were based on SSSD datasets, stemming from mean industry data (Wilson, 2019). The launch campaign was applied using an SSSD dataset consisting of generic data on energy consumption, water consumption and chemicals use for a 21-day launch campaign period, including test firings, directly applied for each launch event.

Now we will outline the total rocket compositions for each constellation. This is important because total emissions are a function of the number of single launches and the rocket type. Given the different number of satellites in each constellation, the number of launches varies, as detailed in Table 3. The most current data is used to estimate the number of launches required for all full phase 1 constellation deployments, with the baseline termed here Scenario 1 (McDowell, 2024). A total of 62.6% of the rockets (149 launches) used to launch the three LEO constellations are within our dataset. Where a rocket is not within the database, we utilize generic representative hydrocarbon-fuel (HYC) and hydrogen-fuel (HYD), which covers 35.7% of rockets (85 launches). Finally, for the 1.7% of LEO launches not yet confirmed (4 launches), we utilize 2 generic HYC and 2 generic HYD, based on a 50/50 split. For the hypothetical GEO operator, we take a similar approach allocating 10 generic HYC and 9 generic HYD.

As detailed in Table 3, this equates to Starlink undertaking 127 launches using Falcon-9. For OneWeb, 14 launches were made using Soyuz-FG, 5 launches via Falcon-9, and 2 launches via LVM3. Amazon's Kuiper has confirmed three agreements for 38 launches via United Launch Alliance's (ULA) Vulcan Centaur rocket, 18 via Arianespace's upcoming Ariane-6 rocket, 27 via its Blue Origin New Glenn rocket, and 3 via Falcon-9 ("Amazon makes historic launch investment to advance Project Kuiper," 2022). As stated above, the remaining 4 launches are treated as consisting of 2 generic HYC and 2 generic HYD. For the hypothetical GEO operator with 19 satellites, we consider one satellite per single launch event amounting to the required 19 launches. Due to the availability of more HYC rockets, we model that 10 launches will be made using a HYC and the other 9 on a HYD rocket.

| Rocket Scenario | Const-ellation | Rocket | Rocket Detail Name | Status | Rocket Fuel Type | No. of Satellites | No. of Launches |
|---|---|---|---|---|---|---|---|
| Scenario 1 | Starlink | Falcon-9 | Falcon-9 | Modeled | Hydrocarbon | 4425 | 127 |
| Scenario 1 | OneWeb | Soyuz-FG | Soyuz-FG | Modeled | Hydrocarbon | 394 | 14 |
| Scenario 1 | OneWeb | Falcon-9 | Falcon-9 | Modeled | Hydrocarbon | 180 | 5 |
| Scenario 1 | OneWeb | HYD | LVM3 | Representative | Hydrogen | 72 | 2 |
| Scenario 1 | Kuiper | HYD | Ariane-6 | Representative | Hydrogen | 648 | 18 |
| Scenario 1 | Kuiper | HYC | New Glenn | Representative | Hydrocarbon | 972 | 27 |
| Scenario 1 | Kuiper | HYC | Vulcan Centaur | Representative | Hydrocarbon | 1368 | 38 |
| Scenario 1 | Kuiper | Falcon-9 | Falcon-9 | Modeled | Hydrocarbon | 108 | 3 |
| Scenario 1 | Kuiper | HYC | Unknown Hydrocarbon | Representative | Hydrocarbon | 72 | 2 |
| Scenario 1 | Kuiper | HYD | Unknown Hydrogen | Representative | Hydrogen | 72 | 2 |
| Scenario 1 | GEO | HYC | Unknown Hydrocarbon | Representative | Hydrocarbon | 10 | 10 |
| Scenario 1 | GEO | HYD | Unknown Hydrogen | Representative | Hydrogen | 9 | 9 |
| Scenario 2 | Starlink | HYD | Unknown Hydrogen | Representative | Hydrogen | 4425 | 127 |
| Scenario 2 | OneWeb | HYD | Unknown Hydrogen | Representative | Hydrogen | 648 | 21 |
| Scenario 2 | Kuiper | HYD | Unknown Hydrogen | Representative | Hydrogen | 3236 | 90 |
| Scenario 2 | GEO | HYD | Unknown Hydrogen | Representative | Hydrogen | 19 | 19 |
| Scenario 3 | Starlink | HYC | Unknown Hydrocarbon | Representative | Hydrocarbon | 4425 | 127 |
| Scenario 3 | OneWeb | HYC | Unknown Hydrocarbon | Representative | Hydrocarbon | 648 | 21 |
| Scenario 3 | Kuiper | HYC | Unknown Hydrocarbon | Representative | Hydrocarbon | 3236 | 90 |
| Scenario 3 | GEO | HYC | Unknown Hydrocarbon | Representative | Hydrocarbon | 19 | 19 |

**SI Table 2|** Rocket compositions. Emissions modeling scenario parameters, including rocket name, fuel type, number of satellites and single event launches for the current, planned and hypothetical constellations.

Additionally, we also explore the impact of a second hypothetical scenario (Scenario 2), where we model all constellations utilizing a HYD fuel-based rocket to launch their full constellations. Consequently, the total launches required in this scenario are 90 (Kuiper), 21 (OneWeb), 127 (Starlink) and 19 (GEO). Finally, in the third scenario (Scenario 3), we model a case where all the constellation operators use HYC fuel-based rockets. See the later section entitled "Results – Rocket Sensitivity", including SI Fig. 4 and 5, to inspect the findings for Scenarios 2 and 3, compared to the Scenario 1 baseline.

## Method - Capacity model

We develop a generalizable model for calculating the total downlink aggregate capacity for Low Earth Orbit (LEO) constellations. This is based on calculating the channel capacity for different satellite altitude and minimum user elevation angles, for three main LEO systems (Kuiper, OneWeb and Starlink), and a comparative hypothetical GEO broadband satellite operator. This is a standard method applied by analysts in the satellite communication industry, as demonstrated in previous studies (Biscarini et al., 2022; Giggenbach et al., 2023; Gongora-Torres et al., 2022; Muttiah, 2023; Popescu, 2017; Saeed et al., 2021). Considering a satellite at an orbital altitude $h$, the user must observe the satellite at a certain minimum elevation angle, $\varepsilon_{min}$ in radians, to connect and access the service. The link distance $d_{km}$ in kilometers, between the user and the satellite is therefore given by equation (1), where $R_E$ is the radius of Earth (6,378 km).

$$d_{km} = R_E \left[ \sqrt{\left(\frac{h + R_E}{R_E}\right)^2 - cos^2\varepsilon_{min}} - sin\varepsilon_{min} \right] \quad (1)$$

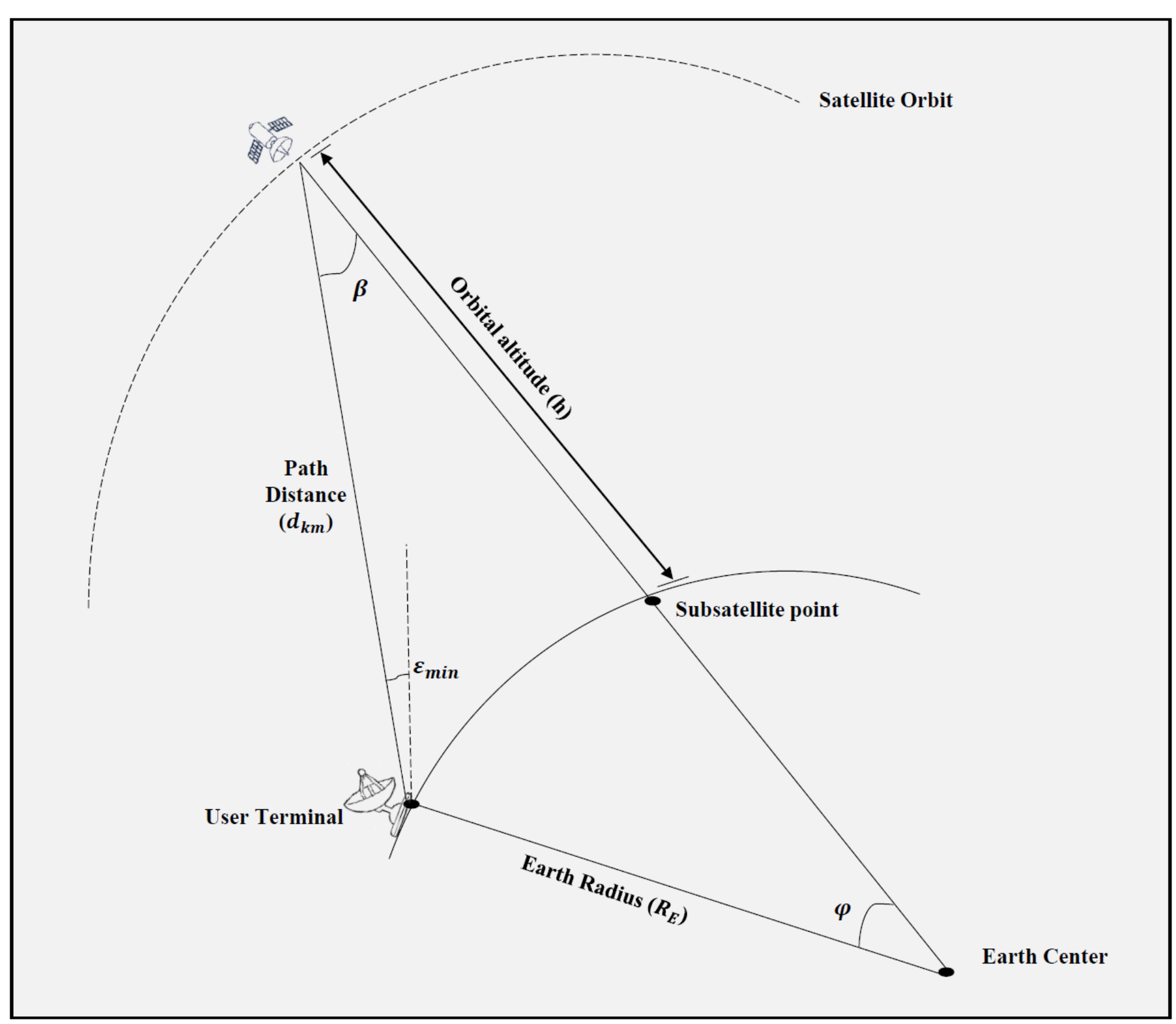

**SI Fig. 3 | Satellite – user terminal orientation.** Illustrating the altitude, minimum elevation angle, satellite and earth centric angles.

The link distance $d_{km}$ determines the amount of free space path loss from the user to the satellite and vice versa. The free space path loss, $FSPL_{dB}$ for a given signal carrier frequency in Gigahertz, $f_{GHz}$ can be computed using equation (2) for a range of $\varepsilon_{min}$ as the satellite moves from the user's horizon to zenith point as shown in SI Fig. 3.

$$FSPL_{dB} = 20\log_{10}(f_{GHz}) + 20\log_{10}(d_{km}) + 92.44 \tag{2}$$

Also, within SI Fig. 3 are the $\beta$ satellite centric angle and $\varphi$ earth centric angle. The three angles are related as shown in equation (3).

$$\varepsilon_{min} + \beta + \varphi = 90 \tag{3}$$

Therefore, the coverage area of a single satellite can be calculated using equation (4).

$$SAT_{coverage} = 2\pi {R_E}^2(1 - \cos\beta) \tag{4}$$

We then estimate the downlink channel capacity of a satellite. First, the energy per bit to noise power spectral density ratio $\left(\frac{E_b}{N_O}\right)_{dB}$ at the user terminal end is given by equation (5).

$$\left(\frac{E_b}{N_O}\right)_{dB} = EIRPD_{sd} + \frac{G}{T} - FSPL_{dB} - L_{All} - 10 \cdot log_{10}(k \cdot T) \tag{5}$$

Where, $EIRPD_{sd}$ is the effective isotropic radiated power density, $\frac{G}{T}$ the receiver antenna gain, $L_{All}$ for all other losses (rain fade, gaseous absorption, polarization, feeder and pointing losses), $k$ the Boltzmann constant, and $T$ the system temperature.

Next, the channel capacity ($C_{channel(Mbps)}$) of the satellite is estimated. LEO systems, as Next Generation High Throughput Satellites (NG-HTS), are capable of Adaptive Coding and Modulation (ACM) (Vasavada et al., 2016). Thus, satellites are able to achieve the best throughput depending on present link conditions. We leverage Modulation and Coding (MODCOD) schemes provided in the DVB-S2 (page 53) (Digital Video Broadcasting Project, n.d.) which is an industry standard for the forward link from satellites to user terminals. The $\left(\frac{E_b}{N_O}\right)_{dB}$ is calculated for different satellite altitude and minimum user elevation angles. Based on the varying calculated $\left(\frac{E_b}{N_O}\right)_{dB}$ and considering a clear sky line-of-sight for Additive White Gaussian Noise (AWGN) channel conditions, the MODCOD and coding rate values utilized are presented in Table 4.

| Energy Per Bit to Noise Power Spectral Density Ratio (dB) | Modulation and Coding (MODCOD) | Coding Rate Ratio | Spectral Efficiency (bps/Hz) |
|---|---|---|---|
| -15.0049 to -2.85 | QPSK | 2/9 | 0.434841 |
| -2.84 to -2.03 | QPSK | 13/45 | 0.567805 |
| -2.02 – 0.22 | QPSK | 9/20 | 0.889135 |
| 0.23 – 1.45 | QPSK | 11/20 | 1.088581 |
| 1.46 – 4.73 | 8APSK | 5/9 | 1.647211 |
| 6.3641 - 6.948 | 8APSK | 36/45 | 1.713601 |
| 6.9747 - 7.8815 | 8PSK | 23/36 | 1.896173 |
| 8.0017 - 8.9955 | 8PSK | 25/36 | 2.062148 |
| 9.0011 - 9.169 | 16APSK | 8/15 | 2.10485 |
| 9.3641 - 9.3782 | 16APSK | 26/45 | 2.281645 |
| 9.3922 - 9.9214 | 16APSK | 3/5 | 2.370043 |
| 9.948 - 10.0285 | 16APSK | 28/45 | 2.458441 |
| 10.0554 - 11.0315 | 16APSK | 23/36 | 2.524739 |
| 11.0612 - 11.5177 | 16APSK | 25/36 | 2.745734 |
| 11.5455 - 12.4488 | 16APSK | 13/18 | 2.856231 |
| 12.7485 - 13.8815 | 16APSK | 7/9 | 3.077225 |
| 14.1665 - 14.4899 | 32APSK | 2/3 | 3.291954 |
| 14.505 - 14.5735 | 32APSK | 32/45 | 3.510192 |
| 15.0011 - 15.169 | 32APSK | 11/45 | 3.620536 |

**SI Table 3 | Spectral efficiency values.** Corresponding spectral efficiency value used to calculate satellite channel capacity for each stochastically determined energy per bit to noise power spectral density ratio ($\left(\frac{E_b}{N_O}\right)_{dB}$). The variation in $\left(\frac{E_b}{N_O}\right)_{dB}$ is due to the changing elevation angle, distance from the satellite to the user and atmospheric losses. A clear sky line-of-sight for Additive White Gaussian Noise (AWGN) channel condition is considered.

The channel capacity ($C_{channel(Mbps)}$) is then estimated, as specified in equation (6), using modulation techniques, bandwidth per channel ($BW_{ch(MHz)}$) (from Table 4) and the corresponding spectral efficiency ($n_{eff(bps/Hz)}$) (from Table 3).

$$C_{channel(Mbps)} = BW_{ch(MHz)} \times n_{eff(bps/Hz)} \tag{6}$$

Furthermore, based on the number of beams ($b$), and number of channels ($ch$) used throughout each constellation, it is possible to estimate the satellite capacity ($C_{Sat(Mbps)}$), as per equation (7).

$$C_{Sat(Mbps)} = C_{channel(Mbps)} \cdot b \cdot ch \tag{7}$$

Only approximately two thirds of the capacity of each constellation is capable of being used at any one time, as many satellites are underutilized when over oceans, or other uninhabited areas (except for some enterprise services across the maritime, aviation, military, oil and gas sectors ("Aerolíneas Argentinas Selects Multi Orbit Inflight Connectivity," 2023; Brennan, 2024)). Therefore, in accordance with the literature, the total usable constellation capacity ($C_{usable(Mbps)}$) is approximated as a product of a coverage factor, $cov_{f(\%)}$ and the total capacity (Steele, 2020), as stated in equation (8) (quantified elsewhere in the literature as the 'useful capacity' (Henry, 2019) or 'sellable capacity' (del Portillo et al., 2019)).

$$C_{usable(Mbps)} = C_{Sat(Mbps)} \times N_{sat} \times \frac{cov_{f(\%)}}{100} \quad (8)$$

Where $N_{sat}$ is the total number of satellites in each constellation. The engineering and orbital parameters used in the capacity model are reported in Table 5.

| Parameter | Unit | Kuiper | OneWeb | Starlink | Hypothetical GEO Operator |
|---|---|---|---|---|---|
| **Constellation name** | - | Kuiper | OneWeb | Starlink | GEO |
| **Operator** | - | Amazon | Eutelsat Group | SpaceX | - |
| **Satellites** | - | 3,236 | 648 | 4,425 | 19 |
| **Satellite mass** | kg | 600 | 150 | 260 | 3,300 |
| **Altitude** | km | 610 | 1,200 | 550 | 35,786 |
| **Elevation angle** | Deg | 35 | 45 | 25 | 5 |
| **Signal path distance** | km | 982* | 1,580* | 1,123* | 41,127* |
| **Satellite centric angle** | Deg | 48.4* | 36.5* | 56.5* | 8.67* |
| **Path loss** | dB | 177* | 177* | 174* | 205* |
| **Earth centric angle** | Deg | 6.61* | 8.48* | 8.45* | 76.3* |
| **Average satellites per launch** | - | 36 | 36 | 23 | 1 |
| **Satellites in orbit (phase 1)** | - | - | >630 | 4,425 | 19 |
| **Downlink frequency** | GHz | 17.7, 18.7, 19.7 | 10.7, 11.7, 12.7 | 10.7, 11.7, 12.7 | 10.5, 11.5, 12.5, 13.5 |
| **Channel bandwidth** | GHz | 0.25 | 0.125 | 0.25 | 0.12 |
| **Antenna diameter** | m | 0.9 | 0.65 | 0.6 | 0.5 |
| **Antenna efficiency** | m | 0.6 | 0.6 | 0.6 | 0.55 |
| **Transmit antenna power** | dBW | 30 | 30 | 30 | 64.2 |
| **Receiver gain** | dBi | 31 | 35 | 30 | 33.4 |
| **Beams** | - | 8 | 16 | 8 | 64 |
| **Channels** | - | 6 | 3 | 6 | 155 |
| **Polarization** | - | 1 | 1 | 1 | 2 |

**SI Table 4|** Engineering and orbital parameters, obtained from FCC filings (*Indicates calculated values).

## Method - Demand model

Scenarios are used in science when a lack of robust information is available. For example, regarding different potential futures, as applied here to possible subscriber scenarios. Explicit low, baseline and high subscriber scenario quantities for each constellation are specified in Table 5. We reach these quantities as follows. Firstly, the number of subscribers that operational LEO constellations have acquired by the final quarter of 2022 forms the low adoption scenario (Hicks, 2022a; Jewett, 2022a). This is 2.5 million for Starlink (Jewett, 2022a) and 0.5 million for OneWeb, whereas we use 1.5 million for Kuiper (yet to launch) and 1.5 million for GEO (Boyle, 2023). In contrast, the higher scenario represents twice this existing number of adopted users, predicated on nearly half of the satellites of the planned constellations having been launched thus far (Sheetz, 2022a). For instance, in the United States alone, Starlink has already deployed 1 million user terminals in anticipation of potentially having more than 4.5 million subscribers (Brodkin, 2020). Kuiper is set to 3.5 million (Boyle, 2023; Kohnstamm, 2023), OneWeb to 1 million (Hicks, 2022b), and GEO to 3.5 million as per press briefings and industry analysis (Daehnick et al., 2022), (Post et al., 2024). Finally, the baseline adoption scenario is set at the midpoint of the lower and upper scenarios. Next, the mean capacity per subscriber ($C_{Mbps^{-sub}}$) is estimated by dividing the total usable constellation capacity ($C_{usable(Mbps)}$) by the number of subscribers for each low, baseline or high scenario, $Sub_{(low,baseline,high)}$ (equation 9).

$$C_{Mbps^{-sub}} = \frac{C_{usable(Mbps)}}{Sub_{(low,baseline,high)}} \quad (9)$$

The monthly traffic ($T_{mon(GB)}$) is estimated per subscriber (in Gigabytes) for each of the constellations via equation (10).

$$T_{mon(GB)} = \frac{C_{Mbps^{-sub}} \times 3600 \times 30}{8000} \quad (10)$$

Thus, the mean capacity per subscriber ($C_{Mbps^{-sub}}$) is converted from seconds into the capacity per hour, by multiplying by 3600 (the number of seconds in an hour). Next, the resulting capacity per hour is converted to a monthly value accounting for 30 days. The result is then divided by 8,000 (accounting for the product of 1,000 and 8, for converting from Megabits Per Second to Gigabytes).

Finally, the mean subscribers per area, $Sub_{Area}$ ($Subscribers^{-km^2}$) can be obtained by utilizing the satellite coverage area ($SAT_{coverage}$ ) as defined in equation (11).

$$Sub_{Area}\ (Subscribers^{-km^2}) = \frac{Sub_{(low,baseline,high)}}{SAT_{coverage}} \quad (11)$$

Method - Cost model

The costs associated with launching and operating satellite constellations are quantified by aggregating the initial capital expenditure ($Capex$) and annually recurring operational expenditure ($Opex$) for a period ($n$) in years (representing the lifetime of each asset). The resulting total cost of ownership ($TCO$) forms the basis of estimating the discounted cost over this time period at a specific discount rate ($r$)("Cost of Capital," n.d.), as per equation (12).

$$TCO = Capex + \sum_{\mathrm{t=0}}^{\mathrm{n}} \frac{Opex}{(1+r)^{n}} \quad (12)$$

Next, both the capex and opex components are estimated using equations (13) and (14) respectively.

$$Capex = C_{Man.} + C_{Lch} + C_{Gst} + C_{Fib.} \quad (13)$$

$$Opex = C_{Reg.} + C_{E} + C_{Staff} + C_{Acq} + C_{Maint} \quad (14)$$

Where for capex, $C_{Man.}$ is the satellite manufacturing, $C_{Lch}$ satellite launch, $C_{Gst}$ ground station, and $C_{Fib}$ fiber infrastructure. And for opex, $C_{Reg}$ is regulatory fees, $C_{E}$ is the energy cost, $C_{Staff}$ is the labor force required to run the network, $C_{Acq.}$ is the customer acquisition cost, and finally, $C_{Maint}$ represents maintenance.

The resulting per subscriber cost for capex ($Cap_{sub}$), opex ($Op_{sub}$), and TCO ($TCO_{sub}$), are obtained as follows in equations (15), (16) and (17).

$$Cap_{sub} = \frac{Capex}{Sub_{(\mathrm{low,baseline,high})}} \quad (15)$$

$$Op_{sub} = \frac{Opex}{Sub_{(\mathrm{low,baseline,high})}} \quad (16)$$

$$TCO_{sub} = \frac{TCO}{Sub_{(\mathrm{low,baseline,high})}} \quad (17)$$

Finally, the SCC ($SC_{carbon}$) is calculated by multiplying the total emissions $Total\ Emissions_{cc(t)}$ by US$ 185 per tonne, as established in (Rennert et al., 2022), and specified in equation (18), for the potential climate change impacts.

$$SC_{carbon} = Total\ Emissions_{cc(t)} \times 185_{\mathrm{US\$}} \quad (18)$$

Now the relevant unit costs will be discussed.

## Constellation unit costs

The constellation unit costs are broadly categorized into capex and opex components, and are specified as follows.

For satellite manufacturing we use publicly available information for GEO and LEO constellations. For a GEO satellite, we use the average unit cost of manufacturing a single satellite to be in the range of US$ 100–300 million as detailed in the literature (Reznik et al., 2020), hence a cost of US$ 300 million is adopted for the hypothetical GEO operator considering the recent effects of inflation. The unit cost can be multiplied by the quantity of satellites in the constellation (e.g., 19) to obtain a total manufacturing cost for the hypothetical GEO operator. For LEO operators, there have been no publicly disclosed manufacturing costs (as this is often proprietary information). Thus, we draw on industry literature applying values of US$ 250,000 for Starlink, and US$ 400,000 for OneWeb and Kuiper (Jonas et al., 2019; OneWeb Holdings Limited, 2022; Wang, 2019). This reflects estimates that the cost of manufacturing LEO communication satellites are below US$ 500,000, while SpaceX has a cost of approximately US$ 250,000. The unit costs are multiplied by the number of satellites in the constellation to obtain the total manufacturing cost.

For satellite launch costs, this quantity depends on the launch vehicle, mass of the payload and the destination orbit. The average cost of launching a single GEO communication satellite using popular launch vehicles, like Ariane-5 and the upcoming Ariane-6, is projected to be in the range of US$ 80–130 million (Duggleby, 2020; Wang, 2023). Consequently, the mean value of US$ 105 million per satellite totalling to US$ 1.995 billion is utilized for the hypothetical GEO operator with 19 satellites. The cost for launching LEO satellites with Falcon-9 is US$ 1,050 per kg (Li et al., 2023). Considering the mass of a single Starlink satellite to be 260 kg and 4,425 satellites in the constellation, the total launch cost is approximately, US$ 1.21 billion. Using the same approach for 648 OneWeb satellites weighing 150 kg, the aggregate satellite launch cost is US$ 102 million. The same can be applied for the 3,236 planned 600 kg Kuiper satellites totalling to US$ 2 billion.

For ground station costs, we estimate this quantity based on existing data from publicly traded GEO satellite operators such as SES (SES S.A, 2022). The number of satellites in the constellation, and the expectation that optical inter-satellite links will be used, is considered in estimating the number and cost of ground stations. For GEO and LEO, we use a value of US$ 0.5 million per ground station (adopted from the SES annual report) (Intelsat Ltd, 2024). Thus, for the hypothetical operator GEO with 8 ground stations, the total cost is US$ 4 million. Moreover, scaling the unit costs to Kuiper's 12 planned ground stations (Kohnstamm, 2023), the total cost amount to US$ 6 million. Similarly, the total ground station cost for OneWeb is US$ 22 million for 44 anticipated ground stations (Swinhoe, 2022), and US$ 75 million for Starlink's projected 150 ground stations (Arevalo, 2023).

For regulatory fees, we consult the US FCC. For example, the fees charged to operators grants them market access and operational authority of satellite systems. The fees are paid upon the operation of the first satellite in the constellation. However, multiple satellites collocated in a similar orbital location are considered as one *system* for calculation purposes. As per the FCC regulatory fee factsheet, US$ 117,580 is charged per GEO satellite system and US$ 12,215 for LEO satellite system annually (Federal Communications Commission, 2023). We treat the LEO orbital planes as unique satellite locations and use four co-locations for the hypothetical GEO operator to calculate the total regulation fee for each system. Thus, the regulation fee for the hypothetical GEO operator is estimated at US$ 0.47 million (4 unique

planes), Kuiper US$ 1.2 million (98 unique planes), OneWeb US$1.2 million (100 unique planes) and Starlink US$ 2.3 million (190 unique planes).

Next, we estimate the annual maintenance costs as 10% of the initial capex for installed terrestrial infrastructure, as is common in the literature (Li et al., 2023; Oughton and Frias, 2018). For satellite constellations, the ground station and associated fiber comprise the primary terrestrial infrastructure. Therefore, the maintenance costs are calculated annually as 10% of the initial total ground station and fiber infrastructure capex costs. For the hypothetical GEO operator, the annual costs amount to US$ 1.05 million and US$ 0.63 million for Kuiper. Similarly, the maintenance cost for OneWeb is US$ 2.31 million and US$ 7.88 million for Starlink annually.

For staff costs, we calculate the total amount based on (i) the number of employees and (ii) the average remuneration per employee, taken from established publicly-traded satellite operators. For instance, the total staff costs for OneWeb's 528 employees in 2022 was US$ 99.5 million translating to US$ 0.19 million per employee (OneWeb Holdings Limited, 2022). Applying the same logic, we estimate the total staff cost for Kuiper's 1,200 employees (Kohnstamm, 2023) to US$ 226 million and US$ 418 million for 2,200 SpaceX employees (20% of the company headcount) working on the Starlink project (Johnson and Roulette, 2018). For the hypothetical GEO operator, we utilize a value of 2,000 employees leading to a total staff cost of US$ 377 million (S.A, 2022).

We treat subscriber acquisition costs for each constellation as a function of the anticipated market coverage and business strategy. For instance, GEO operators often adopt a Business-to-Business (B2B) approach, just as the Eutelsat OneWeb group has indicated their intention to focus their LEO network on serving enterprise customers ("Aerolíneas Argentinas Selects Multi Orbit Inflight Connectivity," 2023). Therefore, subscriber acquisition costs are likely to be lower compared to Starlink and Kuiper that are also providing direct services to individual users via a Business-to-Consumer (B2C) strategy. We use a value of US$ 3.3 million as the subscriber acquisition cost for the hypothetical GEO operator and OneWeb, based on marketing cost values from the SES annual financial statement (SES S.A, 2022) and OneWeb annual report (OneWeb Holdings Limited, 2022). Considering the B2C approach, the number of satellites (global coverage) and possibility of resale agreements, as already experienced in some markets (Fakiya, 2023), we scale Starlink subscriber acquisition costs to US$ 23 million and Kuiper to US$ 16 million.

Next, we estimate the ground station energy costs based on the number of gateway stations that a constellation is likely to have. As per the industry analysis, a typical satellite ground station uses 5 MW of power per year (Energy5, 2023). Going by the recent global average energy consumption for business users of US$ 0.153/KWh (Global Petrol Prices, 2023), the annual energy cost for a single ground station can be estimated to be US$ 770. GEO operators such as Intelsat operate an average of 20 ground stations for 56 satellites (Intelsat Ltd, 2024). Therefore, the hypothetical GEO operator is treated here as having 8 ground stations for 19 satellites, hence the total annual energy cost is estimated at US$ 6,160. Although, Eutelsat OneWeb group has not revealed the number of its ground stations, industry analysis and academic research indicates that it might need 44 ground stations (del Portillo et al., 2019; Swinhoe, 2022), leading to the total annual energy costs of US$ 33,880. As for Kuiper, 12 ground stations (Kohnstamm, 2023) are set to be used to operate the constellation, hence a total annual energy cost of US$ 9,240. Lastly, Starlink is projected to operate 150 ground stations for its constellation leading to total annual energy cost of US$ 115,500.

We estimate the fiber infrastructure cost as a function of distance and the number of ground stations that each satellite operator has. Currently, the average installation cost of 1.6 kilometers (km) of optic fiber cable is US$ 10,000 (Kim, 2022). We set the average distance covered by fiber optic cable in a single ground station to 10 km to the nearest fiber Point of Presence (PoP). Using this value, the fiber infrastructure cost for a single ground station amounts to US$ 62,500. Based on the number of ground stations by each operator, we estimate the total fiber infrastructure costs as US$ 0.625 million for the hypothetical GEO operator, US$ 0.75 million for Kuiper, US$ 2.75 million for OneWeb and US$ 9.375 million for Starlink.

In Table 5, the quantity, unit and total amounts of each of the cost parameters are presented.

## Method - Uncertainty

A proportion of inputs that affect the demand, capacity, and cost estimations are uncertain given a lack of evidence to correctly parameterize these values. Specifically, some capacity parameters including altitude, elevation angle, downlink frequencies, receiver gain and antenna diameter are treated as uncertain parameters to account for variation in the design values. For instance, some variations are expected in the satellite altitude due to atmospheric drag. Even though this is always corrected from the Telemetry, Tracking and Command station, we introduce uncertainty in the filed values to account for altitude loses and general variation. Therefore, the range of values for Kuiper are 604 – 616 km, 1189 – 1201 km for OneWeb, Starlink are 539 – 551 km while the GEO altitude is set to a static value of 35,786 km (Commission, 2020, 2018; Johnson and Roulette, 2018). We also account for different minimum elevation angles as the satellites move from the user's horizon to the Zenith point as illustrated in SI Fig. 3. Using the minimum elevation angles from the filing data, we set a range of 35 – 40° for Kuiper, 35 – 60° for OneWeb, 25 – 40° for Starlink and 5 – 15° for GEO (Commission, 2020, 2018; Johnson and Roulette, 2018). For operational frequency, we use a range of values of allocated bands as sourced from the FCC filings. Therefore, the adopted frequency values for Kuiper are 17.7 – 19.7 GHz, 10.7 – 12.7 GHz for both OneWeb and Starlink and 10.5 – 13.5 GHz for GEO (Commission, 2020, 2018; Johnson and Roulette, 2018).

Due to the uncertainty in the conditions of the atmosphere such as rain, gaseous and cloud attenuation, we vary the overall atmospheric losses from 1 dB in the lower limit to 18 dB in the upper limit. Similarly, the variation in the receiver gain due to fluctuation in the amplifier floor noise levels are modeled by considering a range of values based on the baseline figures filed with FCC. The receiver gain value ranges are 28 – 35 dB (Kuiper), 32 – 38 dB (OneWeb), 27 – 35 dB (Starlink) and 30 – 34 dB (GEO) (Commission, 2020, 2018; Johnson and Roulette, 2018). We also use a range of user antenna diameter values based to account for different designs. The value ranges are deviated around the submitted figure to the FCC and are 0.9 m (Kuiper), 0.5 – 0.65 m (OneWeb and Starlink) and 5 – 10 m (GEO). Notably, the antenna diameter for Kuiper is not varied since it is yet to launch any different design as opposed to existing constellations that have changed their user terminal antenna designs (Davis, 2023).

As already discussed, future demand adoption of satellite broadband services is also unknown. Publicly available subscriber data are utilized to account for the baseline adoption scenario. This includes an expectation of 2.5 million future subscribers for Kuiper (0 currently), 0.8 million for OneWeb (0.2 million currently), 3.5 million for Starlink (2.2 million currently), and 2.5 million for GEO (Hicks, 2022b; Jewett, 2022b; Kohnstamm, 2023; Post et al., 2024; Sheetz, 2022b). All parameter values utilized in the uncertainty analysis are provided in Table 5, with the data sources provided in Table 6.

| Type | Parameter | Kuiper | | | OneWeb | | | Starlink | | | GEO | | | Unit |
|---|---|---|---|---|---|---|---|---|---|---|---|---|---|---|
| | | Low | Baseline | High | Low | Baseline | High | Low | Baseline | High | Low | Baseline | High | |
| Capacity | Altitude | 604 | 610 | 616 | 1189 | 1201 | 1205 | 539 | 550 | 551 | - | 35786 | - | km |
| | Elevation Angle | 35 | 40 | 50 | 45 | 50 | 60 | 25 | 30 | 40 | 5 | 10 | 15 | Deg |
| | Downlink Frequency | 17.7 | 18.7 | 19.7 | 10.7 | 11.7 | 12.7 | 10.7 | 11.7 | 12.7 | 10.5 | 12.5 | 13.5 | GHz |
| | Atmospheric Losses | 1 | 10 | 18 | 1 | 10 | 18 | 1 | 10 | 18 | 1 | 10 | 18 | dB |
| | Receiver Gain | 28 | 31 | 35 | 32 | 35 | 38 | 27 | 30 | 35 | 30 | 33.4 | 34 | dB |
| | Antenna Diameter | 0.9 | 0.9 | 0.9 | 0.5 | 0.5 | 0.65 | 0.5 | 0.6 | 0.65 | 5 | 10 | 10 | m |
| Demand | Subscribers | 1.5 | 2.5 | 3.5 | 0.5 | 0.8 | 1 | 2.5 | 3.5 | 4.5 | 1.5 | 2.5 | 3.5 | mn |
| Cost | Satellite Manufacturing | 0.3 | 0.4 | 0.5 | 0.3 | 0.4 | 0.5 | 0.2 | 0.25 | 0.3 | 200 | 300 | 400 | US$ mn |
| | Satellite Launch | 0.6 | 0.63 | 0.65 | 0.1475 | 0.1575 | 0.1675 | 0.263 | 0.273 | 0.283 | 80 | 105 | 130 | US$ mn |
| | Ground Station | 0.45 | 0.5 | 0.6 | 0.45 | 0.5 | 0.6 | 0.45 | 0.5 | 0.6 | 0.45 | 0.5 | 0.6 | US$ mn |
| | Fiber Infrastructure | 0.03125 | 0.046875 | 0.06 | 0.03125 | 0.046875 | 0.06 | 0.03125 | 0.046875 | 0.06 | 0.03125 | 0.046875 | 0.06 | US$ mn |
| | Regulatory Fees | 0.011215 | 0.012215 | 0.013215 | 0.011215 | 0.012215 | 0.013215 | 0.011215 | 0.012215 | 0.013215 | 0.10758 | 0.11758 | 0.12758 | US$ mn |
| | Ground Station Energy | 600 | 770 | 800 | 600 | 770 | 800 | 600 | 770 | 800 | 600 | 770 | 800 | US$ mn |
| | Staff | 0.1 | 0.188 | 0.25 | 0.1 | 0.188 | 0.25 | 0.1 | 0.188 | 0.25 | 0.1 | 0.188 | 0.25 | US$ mn |
| | Subscriber Acquisition | 15.9 | 16 | 16.1 | 3.2 | 3.3 | 3.4 | 22.9 | 23 | 23.1 | 3.2 | 3.3 | 3.4 | US$ mn |
| | Maintenance | 0.58 | 0.65 | 0.78 | 2.38 | 2.48 | 2.58 | 8.44 | 8.4375 | 8.54 | 0.35 | 0.45 | 0.55 | US$ mn |

**SI Table 5 | Uncertainty inputs.** Capacity, demand and cost parameters for the three constellations.

## Method - Data availability

All datasets utilized are specified in Table 6.

| Model Part | Data Description | Reference |
|---|---|---|
| **Capacity** | Starlink Federal Communications Commission (FCC) Filing | (Federal Communication Commission, 2021) |
| **Capacity** | Kuiper FCC Filing | (Federal Communications Commission, 2020) |
| **Capacity** | OneWeb Filing | (Federal Communications Commission, 2017) |
| **Capacity** | Channel Encoding Documentation | (Digital Video Broadcasting Project, n.d.) |
| **Emissions** | Falcon-9 Technical Details | (SpaceX LLC, 2021) |
| **Emissions** | Soyuz-FG Technical Details | (Zak, n.d.) |
| **Emissions** | Ariane Technical Details | (ArianneSpace Group, n.d.) |
| **Emissions** | Life-Cycle Assessment Dataset | (Wilson et al., 2018) |
| **Demand** | Number of subscribers | (Brodkin, 2020) |
| **Cost** | Cost Estimates | (Duggleby, 2020; Li et al., 2023) |

**SI Table 6|** Data sources for the model.

## Results – Lifecycle Assessment

With regard to Fig. 4 in the main paper, we here continue discussion of additional results. When considering resource depletion for a HYC rocket, Kuiper contributes 503 t, OneWeb 188 t, Starlink 272 t and GEO 74 t, equivalent to a unit of Antimony, Sb (Fig. 4c). However, the resource depletion is lower for the HYD rocket for Kuiper (66 t), OneWeb (7 t) and GEO (30 t) satellites. Additionally, in terms of environmental freshwater ecotoxicity impacts, the launching of all the planned satellites is estimated to result in $120\times10^8$, $51\times10^8$, $126\times10^8$ and $18\times10^8$ equivalent of PAF.m$^3$.day for Kuiper, OneWeb, Starlink and GEO respectively (Fig. 4d) for a HYC rocket vehicle. In comparison, the HYD rocket is estimated to lead to freshwater ecotoxicity values (PAF.m$^3$.day) of $35\times10^8$, $4\times10^8$, $16\times10^8$ for Kuiper, OneWeb and GEO. Finally, we estimate that the launch of all the satellites in each constellation will also result in significant human ecotoxicity impacts, estimated at 1,045 Cases for Kuiper, 360 Cases for OneWeb, 954 Cases for Starlink, and 153 Cases for GEO given a HYC rocket (Fig. 4e). Estimates for the HYD rocket equate to 249, 25 and 112 human ecotoxicity cases for Kuiper, OneWeb and GEO.

## Results – Rocket sensitivity

In SI Fig. 4 we present comparative annual emission results per subscriber, based on the plausible subscriber scenarios outlined in the method. We compare the two alternative rocket scenarios (Scenarios 2 and 3), against the current/planned baseline (Scenario 1). In general, launching satellites using generic HYC rockets results in marginally higher per subscriber emission results compared to generic HYD rockets. For instance, compared to 303±131 kg $CO_2$eq/subscriber, the annual per subscriber emissions for Kuiper falls to 264±115 kg $CO_2$eq/subscriber when purely using a HYD rocket, or rises to 320±139 kg $CO_2$eq/subscriber when for HYC rockets. Similarly, for OneWeb emissions of 274±101 kg $CO_2$eq/subscriber drops to 194±71 kg $CO_2$eq/subscriber when using HYD, or rises to 235±86 kg $CO_2$eq/subscriber when using HYC. Finally, for Starlink compared to emissions of 172±51 kg $CO_2$eq/subscriber (with rocket re-use), the annual per subscriber emissions rise to 250±75 kg $CO_2$eq/subscriber when using a generic HYD rocket, or to 303±90 kg $CO_2$eq/subscriber when using a generic HYC rocket.

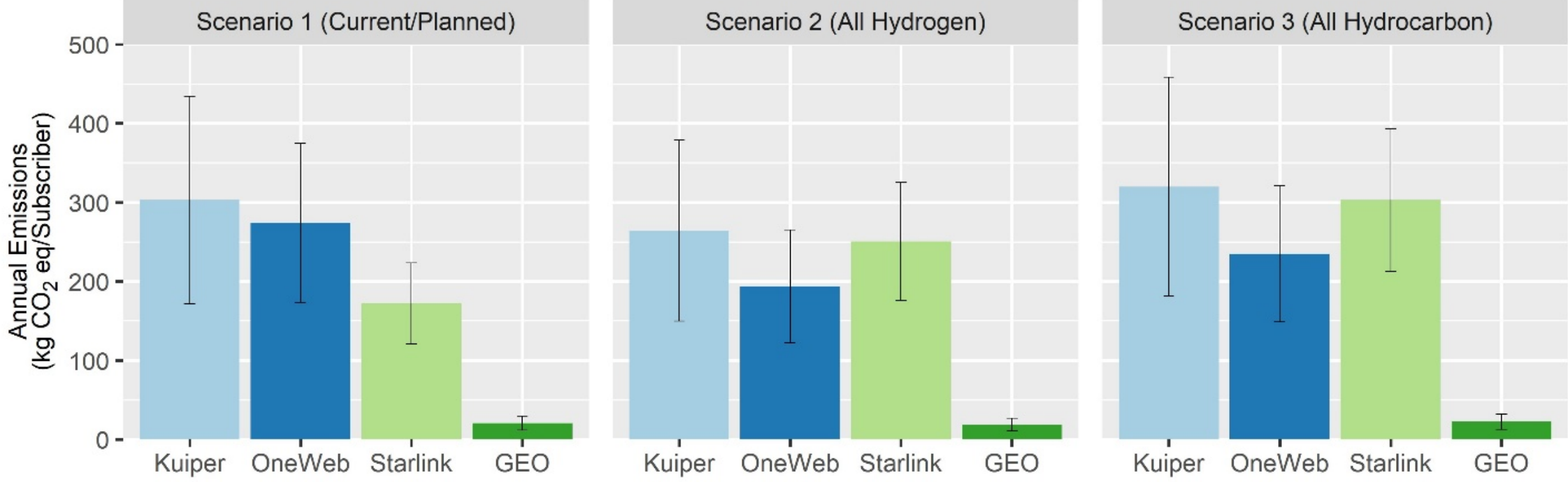

**SI Fig. 4 | Rocket Sensitivity results.** Total annual emissions on a per subscriber basis for rocket Scenarios 1, 2 and 3 (CIs represent the low and high adoption scenarios)

In SI Fig. 5, full total emission results by category for each of the rocket scenarios are presented. The results show mixed trends for different emission categories. For instance, we see climate change emissions between Scenario 2 to 3 increase for Kuiper from 2.93 (HYD) to 3.55 (HYC) Mt $CO_2$eq, for OneWeb from 0.68 (HYD) to 0.83 (HYC) Mt $CO_2$eq, for Starlink from 4.13 (HYD) to 5.01 (HYC) Mt $CO_2$eq, and GEO from 0.62 (HYD) to 0.75 (HYC) Mt $CO_2$eq.

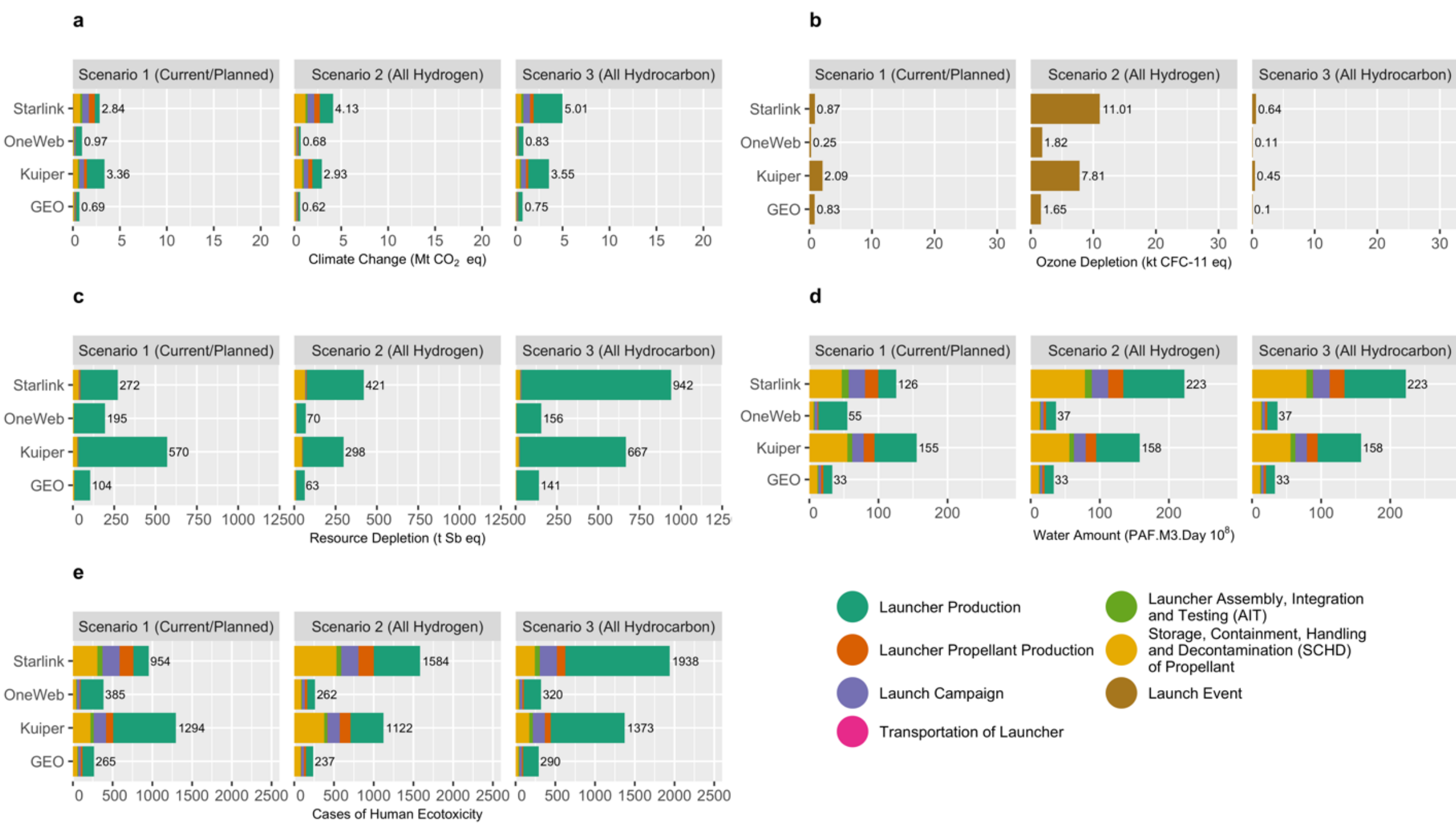

**SI Fig. 5 | Key constellations by environmental impact category. a,** Climate change impacts, **b,** Ozone depletion, **c,** Resource depletion, **d,** Freshwater toxicity, **e,** Human ecotoxicity.

Similarly, for ozone depletion Scenario 2 estimates suggest that HYD rockets result in higher emission impacts (Kuiper, 7.81 kt CFC-11eq, OneWeb 1.82 kt CFC-11eq, Starlink 11.01 kt CFC-11eq and GEO 1.65 kt CFC-11eq) compared to HYC rockets in Scenario 3 (Kuiper, 0.45 kt CFC-11eq, OneWeb 0.11 kt CFC-11eq, Starlink 0.64 kt CFC-11eq and GEO 0.1 kt CFC-11eq). The same trend is seen for other emission categories, with the exception of freshwater toxicity (SI Fig. 5d). In general, HYD rockets have larger climate change and ozone depletion effect, while lower resource depletion impacts and human ecotoxicity cases.

## Results – Constellation capacity

The provided capacity from each LEO constellation has a direct impact on the broadband service each subscriber can access. The design and construction of new infrastructure assets have dramatic sustainability implications (Thacker et al., 2019). Therefore, assessments need to capture the infrastructure trade-off between environmental impacts, and the provided capacity and cost, demonstrating the need for the metrics presented here. Thus, the provided capacity is reported by constellation for different future adoption scenarios, while capturing the stochastic variation in Quality of Service (QoS) resulting from satellite altitude, minimum elevation angle, spectrum bandwidth, antenna and receiver designs.

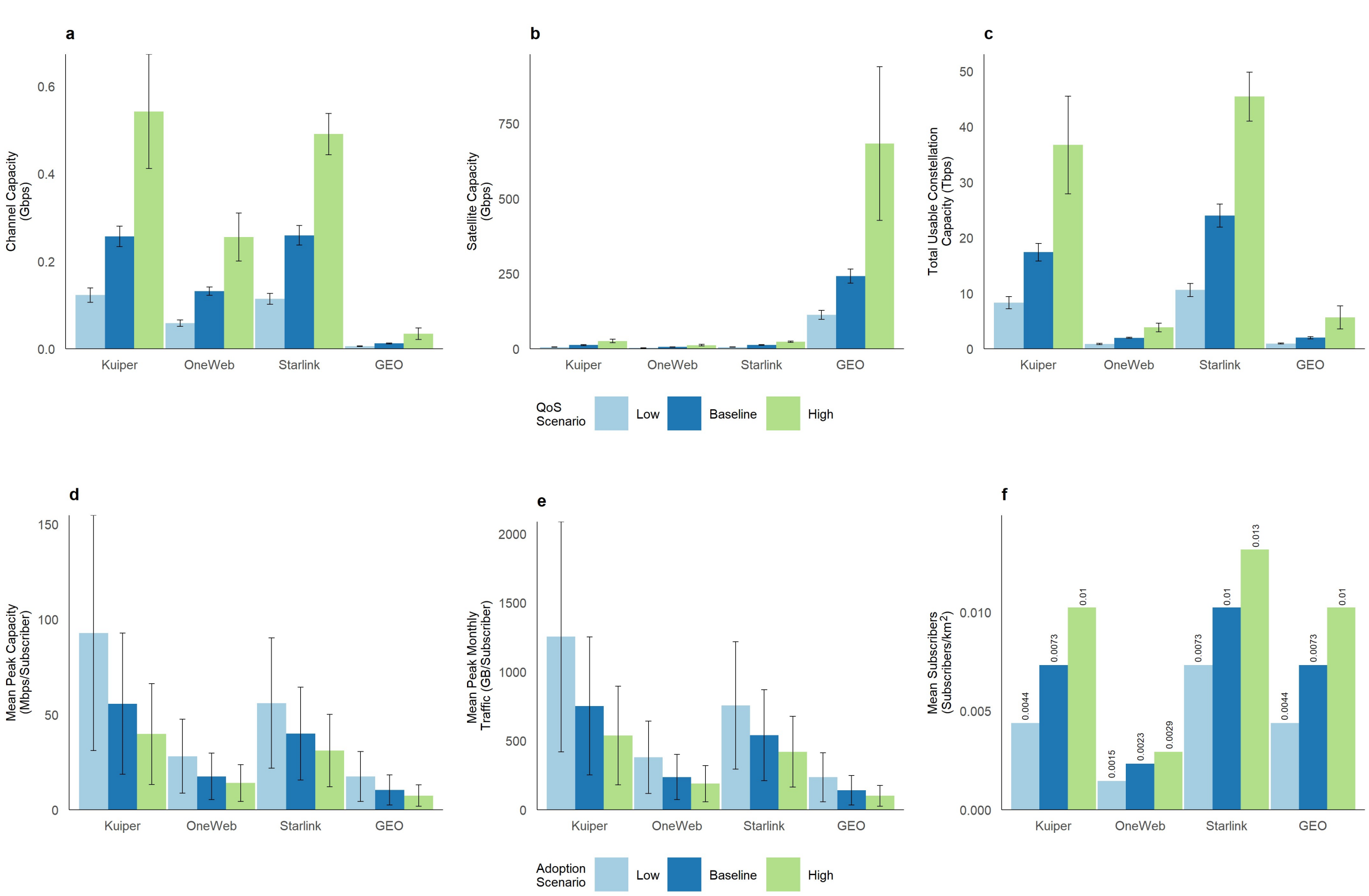

**SI Fig. 6 | Capacity results for different QoS scenarios in a, b, c, d and e (CIs represent uncertainty in QoS due to variation in design parameters including satellite altitude, receiver gain and atmospheric losses at 1 SD. level). a**, Estimated channel capacity, **b**, Estimated single satellite capacity, **c**, Estimated total usable constellation capacity. **d**, Mean peak capacity per subscriber, **e**, Mean peak monthly traffic possible per subscriber, **f**, Mean density of subscribers.

In SI Fig. 6a, the highest baseline channel capacity for a single satellite is estimated at 0.26±0.02 Gbps for Kuiper, which compares to 0.13±0.01 Gbps for OneWeb and 0.26±0.02 Gbps for Starlink. A GEO satellite has the lowest channel capacity at 0.01±0.001 Gbps. The difference in channel capacity is due to the variation in orbital altitude, antenna design and spectrum bandwidth. Overall, with each satellite consisting of a different number of channels, the estimated aggregate satellite capacity is reported in SI Fig. 6b. OneWeb provides the least capacity at 6.33±0.5 Gbps in the baseline QoS scenario, compared to Starlink with 12.45±1.1 Gbps, Kuiper 12.3±1.1 Gbps, and then GEO with the highest at 242±23 Gbps.

As detailed in SI Fig. 6c, the total usable constellation capacity of Starlink is higher than OneWeb and Kuiper (excluding other minor uses including in-flight, off-shore and marine connectivity). For example, Kuiper, OneWeb and Starlink record baseline total usable constellation capacities of 17.3±1.6 Tbps, 1.99±0.1 Tbps and 24±2.1 Tbps. Since, GEO satellites are stationary for a user on Earth, the variation in the total usable constellation capacity is only due to the variation in atmospheric losses. The calculated baseline total usable constellation capacity for the hypothetical GEO operator is 2±0.2 Tbps.

For example, in SI Fig. 6d in the main paper the estimated data rate per subscriber is visualized if all subscribers access the network simultaneously. In the baseline adoption scenario, Kuiper is estimated to provide 56±37 Mbps per subscriber, compared to 18±12 Mbps for OneWeb, 40±24 Mbps per subscriber for Starlink and 10±8 Mbps for a GEO operator during peak mean capacity (all subscribers accessing the network simultaneously). Also, in SI Fig. 6e in the main paper the potential monthly traffic demand capable of being served by Kuiper is estimated at 752±500 GB per subscriber, versus 237±164 GB per subscriber in the case of OneWeb, 540±330 GB per subscriber for Starlink and 141±106 GB per subscriber for a GEO operator.

Finally, for the baseline adoption scenario, Kuiper serves 0.0073 subscribers/$km^2$, compared to 0.0023 $km^2$ for OneWeb, 0.01 $km^2$ for Starlink and 0.0073 $km^2$ GEO as reported in SI Fig. 6f.

## Results – Constellation financial costs

The financial costs to launch each constellation are also estimated, providing insight on required investment (as well as the per subscriber cost), helping to inform strategic choices. Consequently, the hypothetical GEO operator records a capital expenditure (capex) and Total Cost of Ownership (TCO) of US$ 2.7±0.3 billion and US$ 6.2±0.9 billion. Among the LEO operators, Kuiper records the highest overall capex (SI Fig. 7a), operational expenditure (opex) (SI Fig. 7b) and TCO (SI Fig. 7c) consisting of US$ 3.4±0.2 billion, US$ 1±0.2 billion and US$ 4.4±0.3 billion respectively. This is logical as Kuiper has the second largest number of satellites without an in-house launching capability. Starlink is the second most expensive LEO constellation with capex of US$ 2.37±0.27 billion, and estimated TCO of US$ 4.2±0.5 billion, as detailed in SI Fig. 7a, b, and c respectively. Starlink incurs the highest opex of US$ 1.8±0.4 billion. Finally, the lowest cost is estimated for OneWeb with capex, opex and TCO of US$ 0.42±0.04 billion, US$ 1±0.2 billion and US$ 1.42±0.2 billion.

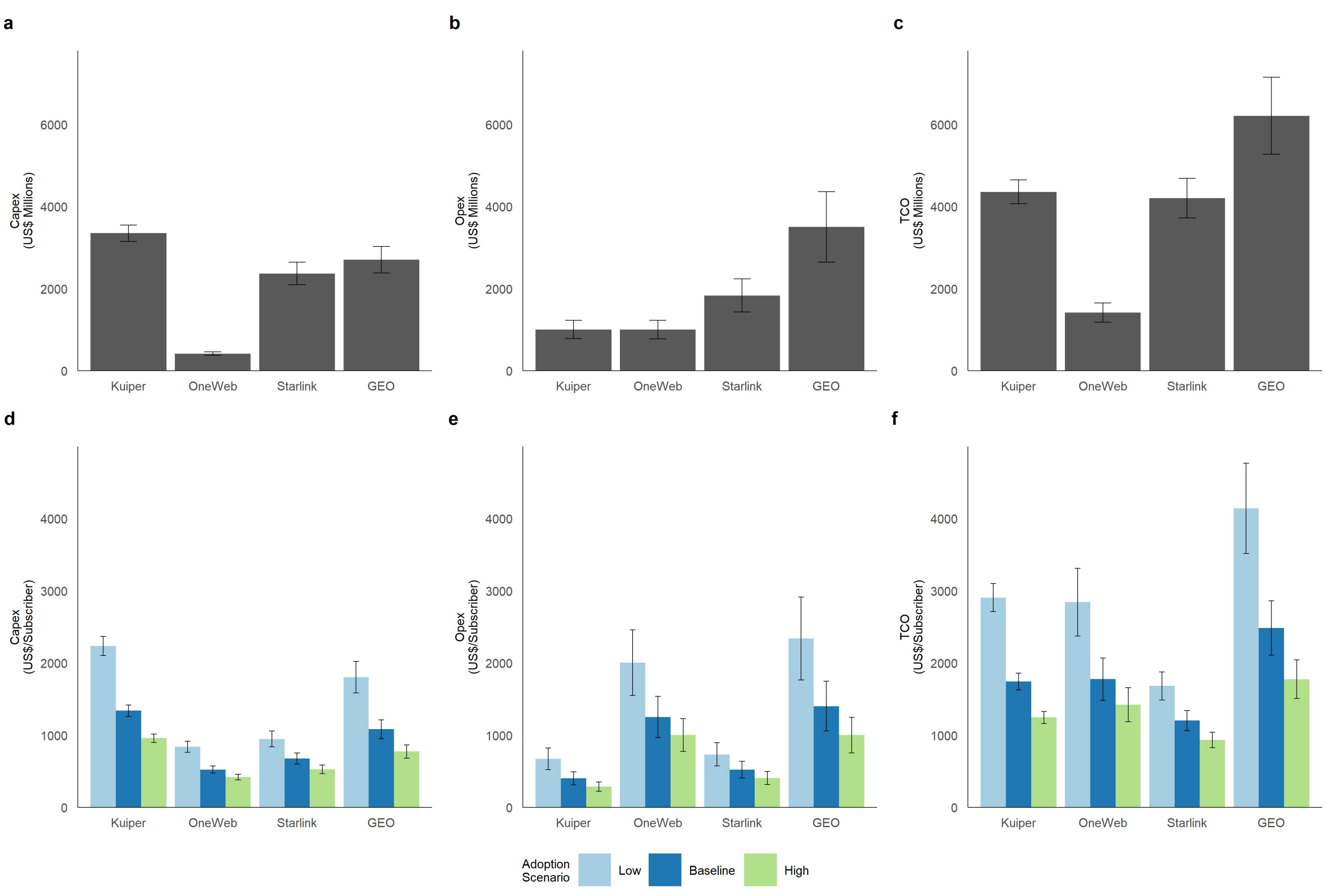

**SI Fig. 7 | Constellation financial costs. a,** Total capex (CIs 1 SD.), **b,** Total opex (CIs 1 SD.), **c,** TCO (CIs 1 SD.), **d,** Capex per subscriber (CIs 1 SD.), **e,** Opex per subscriber (CIs 1 SD.), **f,** TCO per subscriber (CIs 1 SD.).

However, the total financial costs of each constellation do not account for the number of subscribers served, and thus the cost efficiency of each constellation per subscriber. The hypothetical GEO operator has the highest per subscriber cost over its 15-year lifespan of US$ 1,083±131 (capex per subscriber), US$ 1,402±344 (opex per subscriber) and US$ 2,486±376 (TCO per subscriber). For LEO constellations, Starlink has the highest aggregate cost, but also aims to service the largest number of users. This leads to the lowest per subscriber cost with a capex per subscriber of US$ 677±78, opex per subscriber of US$ 524±116, and TCO per subscriber of US$ 1,202±139, as illustrated in SI Fig. 7d, e, and f. In contrast, OneWeb has the lowest aggregate cost for the baseline scenario, but this translates to subscriber capex of US$ 524±49, subscriber opex of US$ 1,252±284 and subscriber TCO of US$ 1,777±293 due to lower targeted adoption numbers. Finally, Kuiper's values are slightly higher than OneWeb, and much higher than Starlink, for example, with a capex per subscriber of US$ 1,341±79 (48% higher than Starlink), an opex per subscriber of US$ 403±90 (66% lower than OneWeb) and a TCO per subscriber of US$ 1,744±117 (31% higher than OneWeb).